# Acceleration and adiabatic expansion of multi-state fluorescence from a nanofocus

N.A. Güsken[1,2], M. Fu[1], M. Zapf[3], M.P. Nielsen[1,4], P. Dichtl[1], R. Röder[3], A.S. Clark[1], S.A. Maier[1,5], C. Ronning[3] and R.F Oulton*[1]

[1]*Department of Physics, Imperial College London, Prince Consort Road, London SW7 2AZ, UK*

[2]*Department of Materials Science and Engineering, Stanford University, Stanford, CA 94305, USA*

[3]*Friedrich-Schiller-Universität Jena, Max-Wien-Platz 1, 07743 Jena, Germany*

[4]*School of Photovoltaics and Renewable Energy Engineering, UNSW Sydney, Kensington, NSW 2052, Australia*

[5] *Nanoinstitut München, Fakultät für Physik, Ludwig-Maximilians-Universität München, Königinstraße 10, 80539 München, Germany*

*\*Corresponding author: r.oulton@imperial.ac.uk*

**Abstract**

**Since Purcell's seminal report 75 years ago, electromagnetic resonators have been used to control light-matter interactions to make brighter radiation sources and unleash unprecedented control over quantum states of light and matter. Indeed, optical resonators such as microcavities and plasmonic nanostructures offer excellent control but only over a limited spectral range. Strategies to tune both emission and the resonator are often required, which preclude the possibility of enhancing multiple transitions simultaneously. In this letter, we report a more than 590-fold radiative emission enhancement across the telecommunications emission band of $Er^{3+}$-ions in silica using a single non-resonant plasmonic waveguide. Our plasmonic waveguide uses a novel reverse nanofocusing approach to efficiently collect emission, making these devices brighter than all non-plasmonic control samples considered. Remarkably, the high broadband Purcell factor allows us to resolve the Stark-split electric dipole transitions, which are typically only observed under cryogenic conditions. Simultaneous Purcell enhancement of multiple quantum states is of interest for photonic quantum networks as well as on-chip data communications.**

The influence of a local electromagnetic environment on light-matter interaction offers a route to realising faster and brighter light sources with associated control over the quantum states of light and matter. Mastering this control is especially important in sensing and quantum technologies[1] when working with low photon numbers and single emitters, such as isolated molecules[2–5], quantum dots[6] and atoms[7,8], which interact particularly weakly with light. In the absence of a structured electromagnetic environment, fluorescence is distributed across a continuum of states, yielding undirected radiation that can be difficult to harness. Consequently, over the past 20 years research has focussed on controlling fluorescence by directing it into isolated electromagnetic cavity modes using the Purcell effect[9].

Optical cavities localize light in both frequency, quantified by the cavity quality factor, $Q$, and space, quantified by a mode volume, $V_m$, to accelerate fluorescence by the Purcell factor, $F_p = 3\lambda^3 Q/(4\pi^2 n^3 V_m)$, where $\lambda$ is the optical wavelength and $n$ is the refractive index. This has proven to be incredibly effective, with recent works demonstrating exquisite control over individual atomic states[7,10,11], but this comes at the price of requiring innovative ways to tune both cavity and emitter as well as the restriction of access to a single electronic transition at a time. A promising alternative is to couple fluorescence into a single highly confined optical waveguide mode[12–14]. Now the mode's confinement area, $A_m$ and group velocity, $v_g$, determine the Purcell factor, $F_p = 3c\lambda^2/(4\pi n^2 v_g A_m)$[15,16], while the continuum of optical states provides tuning-free enhancement across a broad frequency range[17]. This in principle should provide simultaneous access to multiple electronic transitions at distinct energies[18]. Although slow light propagation in photonic crystal waveguides shows excellent Purcell enhancement[7], their bandwidth-delay product limit prohibits broadband operation. Thus, plasmonic waveguides are especially promising as they offer large bandwidth sub-wavelength confinement ($A_m \ll (\lambda/2n)^2$) without excessive dispersion[19,20]. Furthermore, significantly stronger Purcell enhancements are achievable in plasmonic systems compared to their photonic counterparts[12,13,15,21–23], where increases in confinement have been driven by improvements in nanofabrication technology.

One critical attribute of quantum emitters coupled to strongly mode-confining-systems, which has so far remained elusive, is the efficient collection of the enhanced fluorescence, prior to subsequent routing for processing or detection. In this letter, we demonstrate both the acceleration and efficient reverse nanofocusing of the C-band $4f$-shell fluorescence[24] of rare-earth $Er^{3+}$-ions from a nanoplasmonic waveguide[25–27] to a silicon photonic waveguide. Such solid-state emitters are highly desirable[11] as they offer robust, uniform and reproducible quantum states with emission bands that are suitable for long distance quantum photonic networks at telecommunication wavelengths. Unfortunately, their inherently long lifetimes (~10 ms)[28] have hindered practical exploitation for nanophotonics: a single $Er^{3+}$-ion offers only approximately 100 photons per second with ideal collection. Here, we report a strong net enhancement of the $Er^{3+}$-ion fluorescence rate of $\geq 592 \pm 220$ for ions coupled to plasmonic waveguide modes, due to the Purcell effect. Efficient collection of the enhanced fluorescence is achieved by exploiting nanofocusing methods[19], but in reverse. Notably, all of our demonstrated nanofocusing devices show net improvements in fluorescence over non-plasmonic control devices. The extraordinary enhancements of this system are further underpinned by the observation of multiple enhanced Stark-split electric dipole transitions across the $Er^{3+}$-emission band. This demonstrates the capability to strongly dress multiple atomic transitions simultaneously using a one-dimensional mode continuum at room temperature.

# Er³⁺-ion doped reverse nanofocusing waveguide system

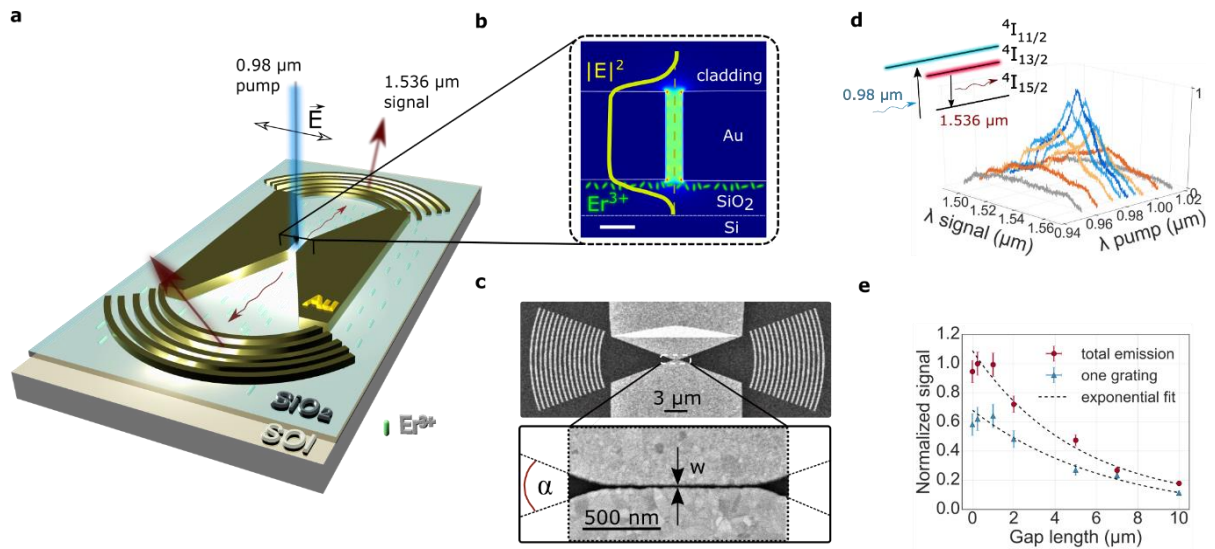

**Figure 1| Er³⁺-ion doped reverse nanofocusing waveguide system**. **a,** Schematic representation of the Er³⁺-ion doped nanofocusing platform pumped at 980 nm with a perpendicular to the gap polarized beam (transverse electric, TE). **b,** Simulated eigenmode of the electric field (field intensity $|E|^2$ in light green) of the waveguide cross-section (50 nm Au, 25 nm SiO$_2$, 220 nm Si, PMMA cladding) at the Er³⁺-ion peak emission wavelength of 1.536 µm. The yellow solid overlay is the field intensity line-cut through the centre of the gap, showing a good overlap of the gap mode with the implanted ions. Er³⁺-ions in green. Scale bar is 20 nm. **c,** Scanning electron microscopy (SEM) top-view of a nanofocusing structure with a waveguide gap length of $L = 1$ µm. The zoom-in discloses a $w = 10$ nm wide metal gap with opening angle $\alpha$. **d,** Measured fluorescence spectra at different signal wavelengths (λ signal) and varius pump wavelengths (λ pump) for a $w = 60\ nm$ wide gap waveguide. The inset schematically illustrates the Er³⁺-energy levels and the fluorescence process involved. **e,** Fluorescence signal as a function of waveguide length at constant gap width $w = 60\ nm$ with two, and only one out-coupling grating, respectively. The signals are normalized to the peak value of the total emission. Dashed lines indicate the exponential signal decay fits.

Figure 1a illustrates a hybrid reverse nanofocusing device, consisting of a plasmonic metal waveguide structure in conjunction with a Si-photonic slab. The central plasmonic waveguide region is optically pumped by a polarized focused laser beam, where only transverse electric (TE) polarized light can excite ions in the gap region. The structure is placed on a silicon-on-insulator (SOI) substrate, where light is free to propagate in the plane confined to a 220 nm thick silicon slab waveguide layer. This device defines a hybrid gap plasmon waveguide (HGPW) system[25,26,29] which is able to efficiently convert light from the plasmonic mode into a photonic mode of the silicon slab. Curved grating couplers[25], which are matched to the circular wavefront of light emitted from the narrow gap waveguide, direct light from the silicon slab waveguide into radiation that can be detected. Er³⁺-ions are implanted in a SiO$_2$ interlayer, which was deposited on top of the SOI substrate prior to the fabrication of the plasmonic waveguide (Supplementary Fig. S1). The implanted ions are distributed over a few nanometers right below the SiO$_2$ surface (Supplementary Fig. S2). Figure 1b, showing the gap mode's electric field distribution, reveals a good overlap with the implanted ions. Ions with electric dipole moments aligned to the gap mode's field distribution and direction experience accelerated fluorescence; hence, emission into this mode will dominate over other slower radiative and nonradiative pathways. Following fluorescence into the plasmonic mode, light subsequently diverges in the taper region. Taper angle and oxide thickness have been specifically engineered to satisfy the eikonal approximation[19,25] enabling experimental gap-to-silicon-slab out-coupling efficiencies of > 80% (Supplementary Fig. S3, Fig. S4b).

Figure 1c shows a scanning electron microscopy (SEM) top-view of the reverse nanofocusing structure with the inset zooming in on a 10 nm wide and 1 µm long metal gap plasmon waveguide. Figure 1d shows the measured fluorescence spectra as a function of pump wavelength for a reverse focusing device. The inset illustrates the excitation pathway from the $4f$-orbital ground level, $^4I_{15/2}$, to the $^4I_{11/2}$ level and the radiative transition from the $^4I_{13/2}$ level back to the ground level. The measurement shows the characteristic spectral shape of Er-ions implanted and activated in $SiO_2$ with a peak signal centred at 1536 nm for a resonant pump wavelength of 980 nm[30].

Figure 1e shows the fluorescence signal for different gap lengths but at constant gap width. The exponential decay, attributed to mode propagation loss, confirms the fluorescence coupling via the waveguide. By using an iris in the image plane we selectively detected light from one grating and the waveguide region alone (second grating blocked). Here, the signal is just over half of the total emission, in which light from both gratings and the waveguide region is collected. This finding in combination with the exponential power loss observed, suggests that most detected fluorescence couples directly into the waveguide and is then directed as radiation from the coupling gratings. The optimal fluorescence, observed for a 1 µm long waveguide, represents a trade-off between propagation length and the number of excited $Er^{3+}$-ions coupled to the plasmonic waveguide. This was the length used in subsequent studies.

## Fluorescence enhancement and the Purcell effect

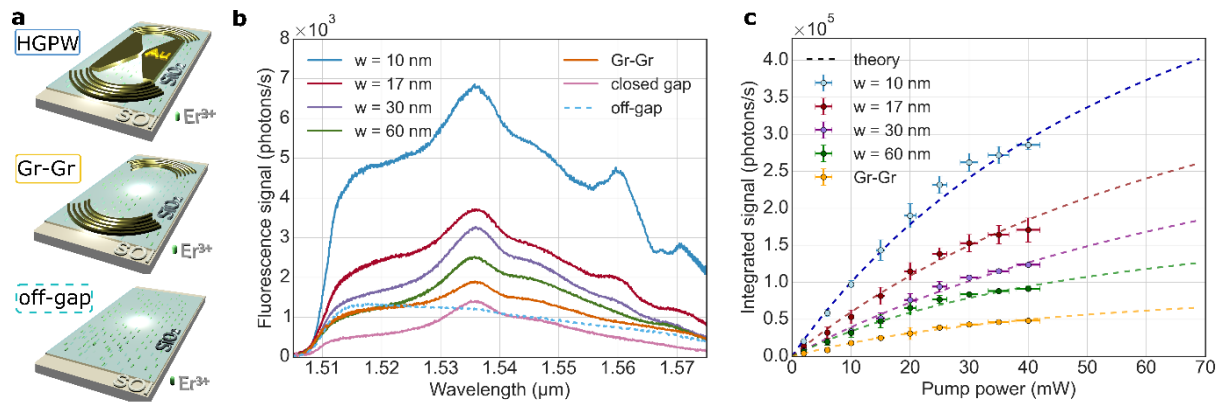

**Figure 2| Fluorescence spectra and power-dependent measurements for different gap widths. a,** Illustration of the compared devices, namely: hybrid gap plasmon waveguide system (HGPW, described in Fig. 1c), only out-coupling gratings (Gr-Gr) and substrate with implanted ions (off-gap), all fabricated on the same sample. Illumination spot in white. **b,** Fluorescence spectra measured for the devices shown in **(a)** with different HGPW gap widths $w$ ($w = 0$ for closed gap) at a pump power of 40 mW and an excitation wavelength of 980 nm in TE polarization (CW pump laser, $\varnothing \approx 2$ µm) for 1 µm long waveguides. **c,** Laser pump power dependence of the integrated spectra from **(b)** with corresponding theoretical fits (Supplementary Eq. (S4.5) and Eq. (S4.8)) based on the three-level rate equation model (Supplementary Information S3 and S4). The extracted values for the saturation signal and power are listed in Supplementary Information Table S1. Note that the excitation pump power was limited to avoid any changes to the device morphology.

Figure 2a shows the three device configurations studied to verify the enhancement capabilities of the nanofocusing waveguide. In the first instance, we consider metal waveguide structures (HGPW) of varying gap width, $w$, to investigate the plasmonic fluorescence enhancement. Secondly, a region of the sample with only out-coupling gratings (Gr-Gr) serves as a control device, to gauge the amount of $Er^{3+}$-fluorescence that couples to the gratings via the silicon slab. Finally, we consider a region of the sample without metallic structures (off-gap), offering a second control measurement for the proportion of $Er^{3+}$-fluorescence radiated directly to free space. In each case, the device is centrally illuminated at normal incidence from the top with a focal spot of $\varnothing \approx 2$ µm at a pump wavelength of 980 nm in TE polarization (perpendicular to the gap) as described above (Fig. 1a). Note that a closed HGPW (i.e. $w = 0$) represents an additional control device (closed gap), to assess the $Er^{3+}$-fluorescence from the triangular taper regions on either side of the HGPW.

Figure 2b shows the corresponding fluorescence spectra for the HGPW devices of varying gap width, and the three control devices: the Gr-Gr, off-gap and closed gap configuration. Importantly, we observe a net emission enhancement for all HGPW devices ($w \neq 0$) compared to the three control devices. The emission enhancement occurs despite the metallized regions restricting the number of $Er^{3+}$-ions that can be excited and also despite any plasmonic propagation or coupling losses. Moreover, the fluorescence continues to improve with reducing gap width, even with the reduction in the number of $Er^{3+}$-ions contributing to the emission. This is remarkable as a substantial number of ions is still being excited in the taper regions, as is evident from the closed gap control device. These observations unambiguously demonstrate a fluorescence rate enhancement of $Er^{3+}$-ions by the HGPW combined with efficient collection enabled by the reverse nanofocusing technique. Note also that the entire emission spectrum of the $Er^{3+}$-ions is enhanced, highlighting the broadband nature of the effect.

Figure 2c shows the pump power dependence of the integrated spectra from Figure 2b. The integrated signal increases with decreasing waveguide width for all powers.

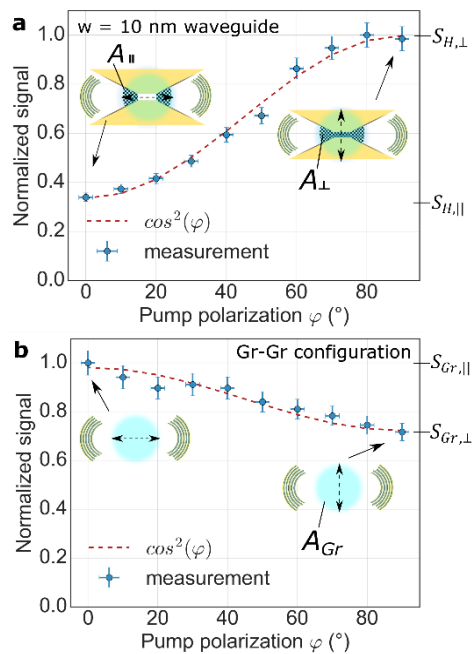

**Figure 3| Dependence of emission on pump beam polarization.** Central illumination with changing polarization angle $\varphi$ at a wavelength of 980 nm for **a,** the HGPW device with a $w = 10\ nm$ gap waveguide and **b,** the Gr-Gr control device. The double arrows indicate the polarization of the pump beam while $A_{//}$ and $A_{\perp}$ in **(a)** refer to the illuminated area in the different polarization configurations (dark shaded area) while $A_{Gr}$ in **(b)** is the focal spot area.

The diminishing number of ions coupling to the HGPW for narrowing gaps clearly dominates the observed signal. We can quantify this effect by using the fact that only pump light polarized perpendicular to the gap may excite the ions below the gap, as verified by simulations (Supplementary Information S5). This allows us to effectively switch the plasmonically enhanced ions on ($\varphi = 90°$) and off ($\varphi = 0°$). Figure 3a, plots the integrated $Er^{3+}$-signal for the 10 nm gap waveguide as a function of pump polarization angle, $\varphi$. Over 78% of the fluorescence comes from plasmonically enhanced $Er^{3+}$-ions within the gap, even though they represent only a small proportion of the total number of fluorescing ions.

Figure 3b indicates a small degree of polarization anisotropy in the Gr-Gr control sample, due to polarization selectivity of the $Er^{3+}$-emission, the slab waveguide and coupler gratings. The enhanced emission thus commands a larger proportion of the total emission, $f = 1 - S_{H,\parallel} S_{Gr,\perp} / S_{Gr,\parallel} S_{H,\perp}$, where $S_{H,j}$, $S_{Gr,j}$ are the signal values in parallel ($j = \parallel$) or perpendicular ($j = \perp$) polarization to the gap of the HGPW and Gr-Gr device, respectively (Fig. 3). For the 10 nm gap waveguide, we find that a fraction $f \geq 78\%$ (Supplementary Information S5) of the collected signal from the HGPW device originates from ions in the gap alone. This occurs despite the small fraction of excited ions situated in the gap.

Figure 2c shows the pump power dependence of the integrated spectra for various gaps and the Gr-Gr configurations. The observed fluorescence enhancement may arise due to both improved excitation as well as accelerated fluorescence. To disentangle the two processes, we consider the saturation behaviour of the $Er^{3+}$-ions. At saturation, the rate of $Er^{3+}$-ion fluorescence is independent of the excitation power and is thus determined only by the fluorescence rate (Supplementary Information S3). The emission from $Er^{3+}$-ions is well described by a three-level rate equation, which forms the basis for a saturation model. Here, our saturation model also accounts for the uniform in-plane distribution of ions across the substrate and the Gaussian excitation beam (Supplementary Information S4). In the experiment, illuminated ions placed at different positions in the gap can contribute to the detected signal differently, as the strength of Purcell effect depends on the position of the emitter within the waveguide. Hence, in order to provide an experimental estimate of the Purcell factor, we identify the ratio of the saturated fluorescence signals in the HGPW device, $S_{Gap,sat}$ and Gr-Gr device, $S_{Gr,sat}$, averaged over the ions which contribute to the signal in the gap wavguide and Gr-Gr device, respectively:

$$\langle F_p \rangle \geq \frac{S_{Gap,sat}}{S_{Gr,sat}} \frac{A_{Gr}}{L \cdot w} \frac{\varepsilon_{c,Gr}}{\varepsilon_{c,Gap}}, \qquad (1)$$

where, $L$ is the waveguide gap length, $w$ is the gap width and $A_{Gr}$ is the area of the illumination beam, as illustrated in Figure 3b. The signal collection efficiencies from the excited regions on the sample are $\varepsilon_{c,Gap}$ for the HGPW device and $\varepsilon_{c,Gr}$ for the Gr-Gr device, estimated from simulations (Supplementary Information S6). $S_{Gap,sat}$ and $S_{Gr,sat}$ are extracted from fitting Supplementary Equation (S4.5) and Supplementary Equation (S4.8) to the measured integrated fluorescence signals as shown in Figure 2c. The parameter values for Equation (1) are listed in Table S1 and S2 for the various waveguide gap widths.

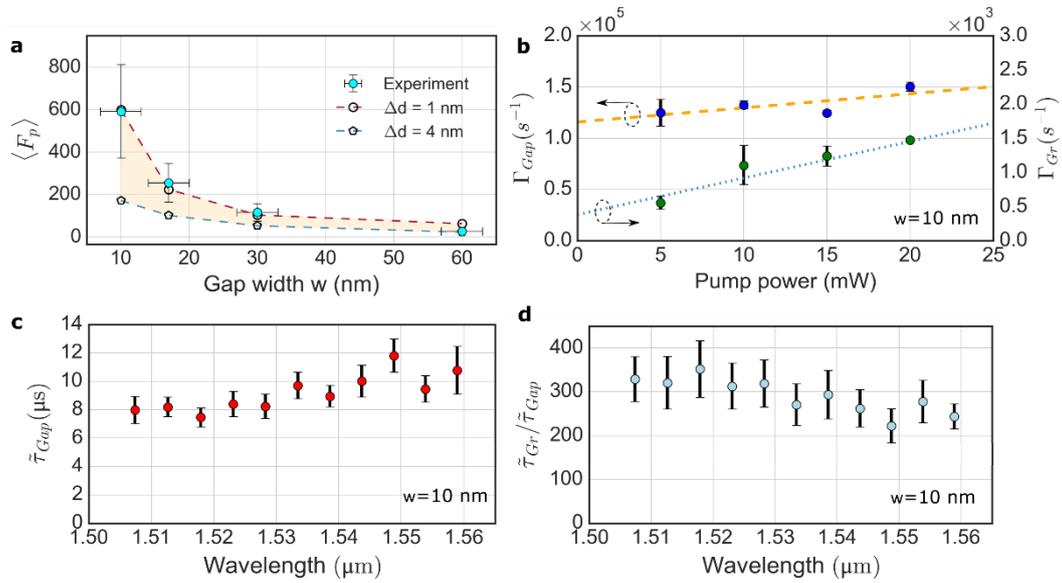

**Figure 4| Purcell factor and emission rate enhancement. a,** Average Purcell factor $\langle F_p \rangle$ in the gap waveguide for various gap widths $w$, extracted from power dependent measurements (Eq. (1)). Dashed lines and circles indicate the simulated average Purcell factor $\langle F_p \rangle$ obtained from averaging over different positions across the gap for a dipole emitter placed at distance $\Delta d = [1\ nm, 4\ nm]$ below the gap waveguide. Here, only dipoles aligned perpendicular to the gap are considered. **b**, Emission rates extracted from time resolved fluorescence spectroscopy comparing the Gr-Gr with the HGPW device ($w = 10\ nm$). The y-intercept yields the emission rates for the gap, $\Gamma_{Gap}$, and the grating device, $\Gamma_{Gr}$. The dotted and dashed lines are the linear fits for each case. **c,** Lifetime wavelength dependence for ions placed in the HGPW ($w = 10\ nm$). **d,** Spectrally resolved lifetime ratio $\tilde{\tau}_{Gr}/\tilde{\tau}_{Gap}$ in the HGPW device ($w = 10\ nm$).

Figure 4 presents the average Purcell factor $\langle F_p \rangle$, the emission rates $\Gamma_{Gap}$ and $\Gamma_{Gr}$ in the gap waveguide and Gr-Gr device, the spectrally resolved ion lifetime $\tilde{\tau}_{Gap}$ and the resulting emission rate enhancement $\tilde{\tau}_{Gr}/\tilde{\tau}_{Gap}$. The values were extracted from two complementary methods: i) power dependent fluorescence saturation measurements (Figure 2c); and ii) time-resolved fluorescence spectroscopy. Figure 4a compares the average Purcell factor, $\langle F_p \rangle$, based on experiment (Eq. (1)) and simulations (Supplementary Information S4), for varying gap widths. The latter is shown for the implantation depth, Δd, of 1 nm and 4 nm. The numerical estimate considers horizontal dipoles aligned perpendicular to the gap, providing the Purcell enhancement factor explicitly for transitions accelerated due to coupling to the TE gap waveguide mode. Note that the pump beam is linearly polarized throughout the experiment and only light in TE polarization excites the ions below the gap (Supplementary Information S5).

The experimental results shown in Figure 4a yield an average Purcell factor of $\langle F_p \rangle \geq 592 \pm 220$ for a 10 nm wide waveguide. The measurement is in good agreement with the numerical estimate of the average Purcell factor for ions implanted at $\Delta d = 1$ nm below the SiO$_2$ surface.

Time-dependent measurements, shown in Figure 4b, provide the emission rates, $\Gamma_{Gap} = \tau_{Gap}^{-1}$, for the $w = 10$ nm HGPW and, $\Gamma_{Gr} = \tau_{Gr}^{-1}$, for a Gr-Gr device. In order to extract the natural emission lifetime, these power dependent data were extrapolated to zero pump power. The lifetimes are $\tau_{Gap} = 8.6$ μs $\pm$ 0.8 μs for the ions in the gap of the HGPW device and $\tau_{Gr} = 2.629$ ms $\pm$ 0.667 ms for the Gr-Gr device, which result in an emission rate enhancement of $\Gamma_{Gap}/\Gamma_{Gr} = \tau_{Gr}/\tau_{Gap} \approx 305 \pm 48$.

The Purcell factor is defined as $\langle F_p \rangle = \gamma_R/\gamma_{0,R}$, where here $\gamma_R$ is the radiative decay rate of ions within the gap and $\gamma_{0,R}$ in the radiative rate of the Gr-Gr device. The total emission rates of ions in the gap, $\Gamma_{Gap} = \gamma_R + \gamma_{NR}$, and in the Gr-Gr configuration, $\Gamma_{Gr} = \gamma_{0,R} + \gamma_{0,NR}$, contain both radiative and non-radiative contributions, but their ratio is also related to the Purcell factor by: $\Gamma_{Gap}/\Gamma_{Gr} = \langle F_p \rangle \eta_{q,Gr}/\eta_{q,Gap}$, where $\eta_{q,Gr}$ and $\eta_{q,Gap}$ are quantum efficiencies of the ions in the gap waveguide and the Gr-Gr control device, respectively. Based on this result and the determined $\langle F_p \rangle$, the ratio of internal quantum efficiencies for Er$^{3+}$-ions in the HGPW ($w = 10$ nm) and Gr-Gr devices is $\eta_{q,Gap}/\eta_{q,Gr} = \langle F_p \rangle \Gamma_{Gr}/\Gamma_{Gap} \approx$ 194 % $\pm$ 73%. This confirms a net improvement in quantum efficiency with radiative Purcell enhancement dominating over emission quenching in the gap waveguide device.

While the lifetime values in Figure 4b are based on integration over the entire erbium emission band, we also spectrally resolved the lifetime to show a range of 8 μs – 10 μs in Figure 4c. With the natural lifetime of Er$^{3+}$-ions implanted into a SiO$_2$ host matrix being tens of ms[30,31], our measurements suggest a Purcell enhancement approaching three orders of magnitude in the $w = 10$ nm HGPW device over the full measured spectral range. This coincides with the average radiative Purcell enhancement estimate $\langle F_p \rangle$ shown in Figure 4a. A direct comparison of lifetimes between the HGPW and the Gr-Gr device, shown in Figure 4d, yields a broadband mean fluorescence enhancement factor of $\tilde{\tau}_{Gr}/\tilde{\tau}_{Gap} \approx 260$ across the C-band. Importantly, the device is not restricted by any resonance condition, in stark contrast to conventional cavity-based fluorescence enhancement devices[32,33].

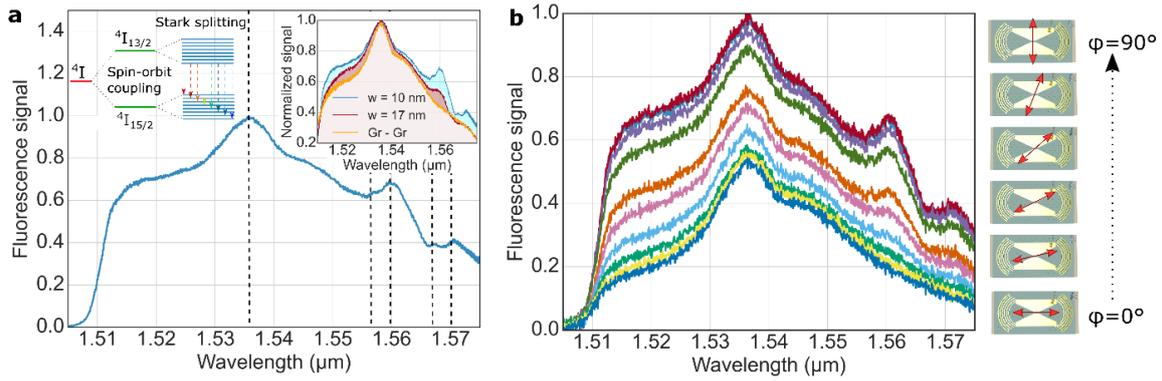

**Figure 5 | Stark splitting of the $^4I_{13/2} \rightarrow {}^4I_{15/2}$ transition levels in Er³⁺: SiO₂ and modal polarization dependence. a,** Measured normalized fluorescence spectra for a $w = 10\ nm$ HGPW at $40\ mW$ CW pump power. Dotted lines mark the energy transitions observed in the experiment. The top left inset schematically illustrates the Stark splitting and the resulting transitions between the $^4I_{13/2}$ and $^4I_{15/2}$ energy levels. The top right inset compares the normalized spectra of the $w = 10\ nm$ HGPW (blue) with a $w = 17\ nm$ HGPW (red) and the signal measured in the Gr-Gr (orange) device. **b,** Polarization dependent fluorescence spectrum for the same settings as shown in **(a)**. The schematics on the right illustrate the polarization angle φ of the pump beam with respect to the gap waveguide orientation. The fluorescence increases monotonously with increasing angle.

At last, we turn the reader's attention to the spectral shape of the Er³⁺-ion fluorescence for the smallest waveguide gap widths. Figure 5a shows the fluorescence spectrum of a $w = 10$ nm wide HGPW device. Here, we observe not only a strong fluorescence enhancement of the central peak at 1.536 µm, but also transition signatures at 1.557 µm, 1.560 µm, 1.567 µm and 1.571 µm. The top right inset in Figure 5a shows the normalized fluorescence spectra for $w = 10$ nm, $w = 17$ nm and a Gr-Gr control device, confirming the change in the fluorescence spectrum's shape with gap width. The acceleration of specific dipole transitions (spectral peaks) can be linked to the Purcell effect as follows. The additional transitions of the Er³⁺-ions embedded in a crystal host matrix result from the Stark splitting of the $^4I_{13/2}$ and $^4I_{15/2}$ energy levels[24,34,35], introduced by local electrical fields in the SiO₂ host matrix at the Er³⁺-sites. At room temperature, these are typically not resolved, and merely broaden the spectrum around its central emission peak[30]. The fact that these signatures are now resolved at room temperature can be explained by the strengthened coupling of the Er³⁺-ion electric dipole (ED) transitions to the plasmonic gap mode, while magnetic dipole (MD) transitions are not enhanced. The smaller the gap width, the smaller the effective mode area and the more energy is strongly confined within the TE gap mode (larger Purcell factor). This results in a large electric local density of states to which light, emitted during ED transitions, couples best. Note, that the transition peaks observed in Figure 5a (dashed lines) can be identified at cryogenic temperatures[36]. Simulations (Supplementary Information S7) confirm that the MD emission enhancement by the plasmonic gap is negligible with modal gap coupling effectively suppressed. Hence, we identify the transitions at 1.557 µm, 1.560 µm, 1.567 µm and 1.571 µm, highlighted in Figure 5a, as ED transitions of trivalent erbium ions embedded in the SiO$_2$ host matrix. Note that the exact energy of the individual ED transition depends strongly on environmental parameters which can be studied in the future.

Figure 5b shows pump beam polarization dependent measurements for the $w = 10$ nm HGPW device. The additional transition peaks are visible only for a pump polarization perpendicular to the gap, where the plasmonic enhancement is enabled (c.f. Fig. 3). Transitioning perpendicular to parallel polarization clearly turns off the additional peaks with the plasmonic enhancement. The ability to selectively access and control multiple ED transitions in this way over a broad frequency range at room temperature via a single waveguide platform is of interest for a wide range of materials and technologies[37]. The facile combination with systems, such as emitters embedded in layers of 2D materials[38], is particularly promising, for which our approach reduces the need for additional methods to tune specific emitter transitions and localized modes.

In summary, we have presented an efficient reverse nanofocusing waveguide platform, which exploits the Purcell effect to locally enhance the fluorescence rate of $Er^{3+}$-ions across the entire optical communications C-band. Boosting the interaction between light and solid state emitters in a controlled fashion gave rise to brighter and faster light sources, with all plasmonic devices offering net enhancement over non-plasmonic control devices. The hybrid plasmonic gap waveguide supports a $592 \pm 220$-fold radiative Purcell enhancement while direct lifetime measurements revealed a total emission rate enhancement of $305 \pm 48$ across the $Er^{3+}$-spectrum. The measurements suggest an emission rate acceleration approaching three orders of magnitude compared to $Er^{3+}$-ions in silica. The polarization-selective emission of $Er^{3+}$-ions placed beneath the gap region allowed switching of the plasmonically enhanced fluorescence on and off, revealing the clear dominance of the small number of enhanced gap ions to the overall fluorescence signal. The hybrid waveguide platform also enabled the enhancement of the Stark-split electric dipole transitions of the $Er^{3+}$-fluorescence band at room temperature, showcasing the ability to access and selectively enhance multiple electric dipole transitions by a single plasmonic waveguide mode despite phonon broadening. Our non-resonant plasmonic waveguide platform allows for unprecedented directionality and control over the enhanced fluorescence signal, while relying on facile and Si-photonics compatible fabrication. This control will enable the development of quantum technologies using single emitters, such as single photon sources[7,39], and ensembles of ions, such as quantum memories[40]. Our approach does not limit the choice of emitter materials, opening up a range of possible frequency regimes and applications for fluorescence enhancement such as on-chip communication and sensing. The possibility to directly contact and control the waveguide electrically provides a simple method to manipulate transition states via the Stark effect offering great potential for opto-electrical signal modulation of quantum emitters.

## Methods

**Sample Fabrication.** An outline of the individual fabrication steps of the reverse nanofocusing hybrid gap plasmonic waveguides with implanted $Er^{3+}$-ions is illustrated in Figure S1 while top-view SEM images of the resulting structures are depicted in Figure S1b-e. First, a 25 nm $SiO_2$ layer was sputtered onto commercially available SOI wafers. After that, Erbium-ions were implanted into the $SiO_2$ interlayer at room temperature using ion implantation with an acceleration voltage of 10 keV, a sample tilting angle of 45° and an ion fluence of $\rho = 1 \times 10^{15}$ cm$^{-2}$. This resulted in a mean ion implantation depth of $(6.0 \pm 2.5)$ nm, as simulated with the Monte-Carlo software package TRIM[41–43]. This is followed by an activation step in which the samples were annealed under controlled atmosphere for one hour at 900 °C. Here, directional ion migration during the annealing process towards the $SiO_2$ interlayer surface can take place (estimated to $\langle y \rangle \approx 3.8$ nm), shifting the ion implantation density peak closer to the surface (Supplementary Information 1). After annealing, hybrid gap plasmonic waveguides (HGPW) were fabricated on top of the $Er^{3+}$-doped $SiO_2$ interlayer. To produce narrow metal-insulator-metal (MIM) gaps, a two-step electron beam lithography (EBL) process with a large range of varying target gap widths has been used resulting in gaps as small as 10 nm. After each EBL step, a 50 nm Au layer was deposited via evaporation at ambient temperature and at a pressure $< 5 \times 10^{-7}$ Torr. Finally, the samples were covered by a thick cladding layer of poly(methyl methacrylate) (PMMA).

**Optical measurements.** The optical characterisation was performed using the setup illustrated in Supplementary Information S8). A CW laser diode at a wavelength of 980 nm was used to pump the erbium ions in the metal gap waveguides. The angle of linear polarization of the incident light was controlled using a λ/2 - waveplate and a polarizer. Here mostly perpendicular to the gap polarized light was used (TE) polarization) to excite the ions below the gap most effectively. The pump beam was focused onto the centre of the plasmonic gap by an infinity-corrected achromatic near-IR objective ($\text{NA} = 0.4, 20x, \text{WD} = 2$ cm). The exact waveguide position with respect to the beam was controlled via a closed loop piezo stage with an IR camera used to image the sample and spot positions. The measurement was performed in reflection and the erbium ion signal was collected via the same objective after coupling to free space via curved grating couplers which were optimized for a centre wavelength of 1.536 µm and TE polarized emission. The collected and collimated erbium signal was analysed using a nitrogen cooled ($-120$ °C) spectrometer (Princeton Acton sp2300). The spectrometer grating used had a groove density of $600$ gr/mm and a blaze wavelength of $1.6$ µm with an efficiency of $\approx 85\%$ at 1536 nm. To select specific regions of the sample, the collected light was focused onto an image plane in which an adjustable iris was positioned. This allowed the signal from only one coupling grating to be collected. However, the main measurements were performed without the pinhole in order to collect the maximum amount of light from the sample via both gratings. Polarization filters were employed to control polarization and power, while suitable short and long pass filters, as shown in Figure S11, were used to ensure that no potential additional pump signal reached the detector. All measurements were performed in the dark at a maximum integration time of 120 seconds.

A second, independent estimate of the fluorescence enhancement in the waveguide system was provided based on time resolved spectroscopy. Here, a modulated pump and detection technique as described in further detail in the Supplementary Information 9, using a 980 nm CW pump laser was employed. The incoming pump as well as the outgoing signal beam are modulated by two phase-locked, out of phase (π-shift) mechanical choppers in order to single out the signal decay in time as illustrated in Supplementary Information Figure S12. Advantages of this technique are that it allows to i) eliminate power dependent effects common to ultrafast pulsed techniques and ii) spectrally resolve the emission rate enhancement over a broad wavelength range.

**Simulations.** Finite-difference-time-domain (3D FDTD) simulations were used to calculate the electric fields within the nanogap waveguide (c.f. Fig. 1), optimize in- and out-coupling efficiency, determine the polarization dependence of incident light as well as the numerical Purcell enhancement estimate. For the latter, the $Er^{3+}$-ion was modelled as an electrical dipole emitter with a perpendicular orientation of the dipole moment to the gap (TE), at a wavelength of 980 nm. The emitted power was determined by a transmission box and a homogeneous mesh of sub-nanometer mesh cell size. The estimate for the Purcell enhancement resulted from dividing the total emission of a dipole placed below the gap waveguide for various gap widths by the transmission resulting from an identical simulation setting in a reference simulation without Au waveguide structures (Gr-Gr device). The magnetic dipole in Figure S10 was simulated in the same manner. In order to estimate the vertical radiation in HGPW and Gr-Gr configuration towards the objective as well as the power coupled into the Si slab from a horizontal dipole, the transmitted power in each direction was calculated and divided by the total emitted power in all direction from an electric dipole aligned parallel to the pump beam polarization in analogy to the experiment.

**Acknowledgement**

The authors thank Anita Chandran and Robert T. Murray from Imperial College London for providing valuable laser equipment and support. We like to thank the Deutsche Forschungsgemeinschaft (DFG) for financial support within the frame of the collaborative research center CRC 1375 "Nonlinear optics down to atomic scales (NOA)", project C5. S.A.M. additionally acknowledges the Lee-Lucas Chair in Physics. N.A.G likes to thank the German National Academy of Sciences Leopoldina for their support via the Leopoldina Postdoc Fellowship (LPDS2020-12).

**Author contributions**

R.F.O., C.R. and N.A.G. conceived and designed the experiments. N.A.G. fabricated and characterized the samples. N.A.G. performed the experiments and analyzed the data. M.F. performed and analyzed the time-resolved spectroscopy. N.A.G. performed the simulations. M.Z. performed the ion-implantation and implantation depth calculations. M.P.N. and P.D. assisted with preliminary characterizations of samples. N.A.G. and R.F.O. wrote the manuscript with input from M.F., M.Z.,M.P.N., P.D., R.R., A.S.C., S.A.M., and C.R.. R.F.O. and S.A.M supervised the project.

**Additional Information**

**Competing interest**

The authors declare no competing financial interest.

# Supplementary information: Acceleration and adiabatic expansion of multi-state fluorescence from a nanofocus

N.A. Güsken[1,2], M. Fu[1], M. Zapf[3], M.P. Nielsen[1,4], P. Dichtl[1], R. Röder[3], A.S. Clark[1], S.A. Maier[1,5], C. Ronning[3] and R.F Oulton*[1]

[1]*Department of Physics, Imperial College London, Prince Consort Road, London SW7 2AZ, UK*

[2]*Department of Materials Science and Engineering, Stanford University, Stanford, CA 94305, USA*

[3]*Friedrich-Schiller-Universität Jena, Max-Wien-Platz 1, 07743 Jena, Germany*

[4]*School of Photovoltaics and Renewable Energy Engineering, UNSW Sydney, Kensington, NSW 2052, Australia*

[5] *Nanoinstitut München, Fakultät für Physik, Ludwig-Maximilians-Universität München, Königinstraße 10, 80539 München, Germany*

## 1. Contents

## 2. Device fabrication

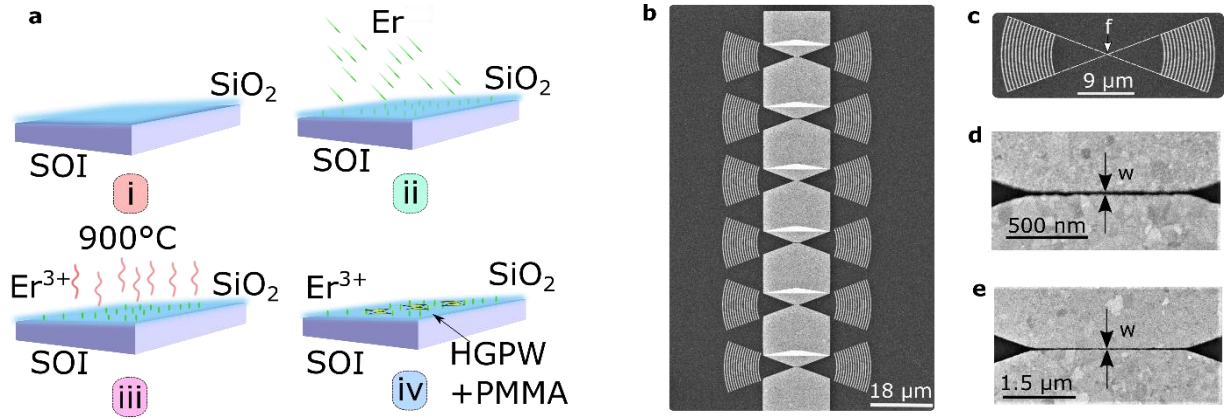

**Figure S1| Fabrication and top views of the erbium doped hybrid gap plasmonic waveguides. a,** i)-iv) Illustration of the individual fabrication steps of the erbium doped HGPW reverse nanofocusing platform. **b,** Array of gold waveguides with different gap widths and focusing gratings on both sides. **c,** Image of the "grating-to-grating" (Gr-Gr) device explained in the main text. The white dotted line overlay highlights the focal position $f$. **d,** 1 μm long gap waveguide with a gap width of $w = 10 \pm 3\ nm$. **e,** 3 μm long gap waveguide with a gap width of $w = 12 \pm 3\ nm$.

An outline of the individual fabrication steps of the reverse nanofocusing hybrid gap plasmonic waveguides (HGPW) with implanted $Er^{3+}$-ions is illustrated in Figure S1a while top view SEM images of the resulting structures are depicted in Figure S1b – e. First, a 25 nm $SiO_2$ layer was deposited onto SOI wafers. Then, erbium ions were implanted into the $SiO_2$ interlayer using ion implantation with an acceleration voltage of 10 keV, a sample tilting angle of 45° and a fluence of $\rho = 1 \times 10^{15}\ cm^{-2}$. This resulted in a mean ion implantation depth of $(6.0 \pm 2.5)$ nm, as simulated with the Monte-Carlo software package TRIM[1,2] and considering sputtering effects, as shown in Figure S2. After that, the samples were annealed under controlled atmosphere for one hour at 900 °C. Subsequent annealing primarily targets defect removal and optical activation of the implanted centres; however, can also cause diffusion of the implanted atoms. Here, we estimate a maximum mean diffusion length of about $\langle y \rangle = 2\sqrt{D\,t} = 3.8$ nm, based on a diffusion coefficient of $D = 10^{-17} cm^2 s^{-1}$ at 900°C for Er in $SiO_2$[3]. Typically this broadens the depth distribution; however, a preferential diffusion towards the surface can also occur, due to the large density of implantation defect states in the penetrated layer compared to the underlying $SiO_2$[4]. This effect results into a re-distribution of the implanted ions and a profile, which is even closer to the surface than the simulated TRIM profile (Fig. S2). After annealing, which in any case optically activates erbium and results in $Er^{3+}$-ions, hybrid gap plasmon waveguides (HGPW) were patterned on top of the $Er^{3+}$-doped $SiO_2$ interlayer. To produce narrow metal-insulator-metal (MIM) gaps, a two-step EBL process was used, resulting in gaps as small as 10 nm. After each EBL step, a 50 nm Au layer was deposited via thermal evaporation at a pressure $< 5 \times 10^{-7}$ Torr. Finally, the samples were covered with a thick poly(methyl methacrylate) (PMMA) cladding layer.

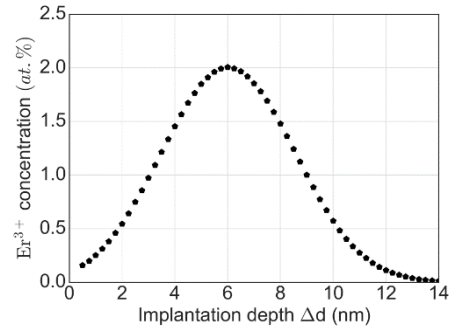

**Figure S2| Implantation depth profile of Er³⁺- ions in a SiO₂ interlayer (before annealing).**

## 3. Hybrid plasmonic gap waveguide: Propagation length and coupling efficiency

In this section, we experimentally and theoretically show that our optimized reverse nanofocusing platform provides coupling efficiencies between gap mode and silicon slab mode of about 80% and propagation lengths of several μm in the most confined state. The system is identical to the HGPW system as described in the main text but without implanted ions. Additionally, grating pitch and duty cycle for the measurement in the main manuscript have been optimized for a centre wavelength of 1536 nm.

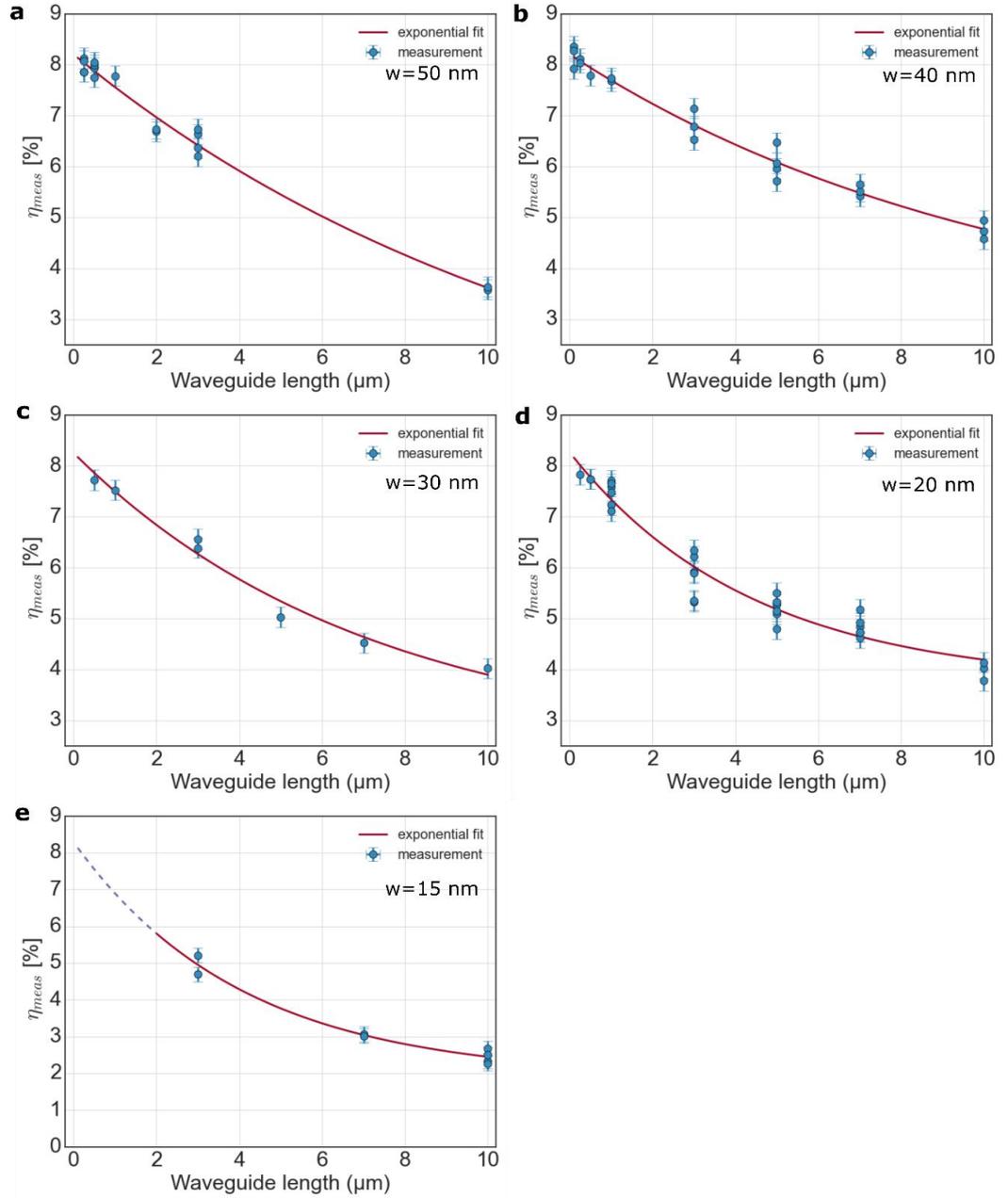

**Figure S3| HGPW cut-back measurements and coupling efficiencies** (Gr-waveguide-Gr) for waveguide widths $w$ of **(a)** $w = 50\ nm$, **(b)** $w = 40\ nm$, **(c)** $w = 30\ nm$, **(d)** $w = 20\ nm$ and **(e)** $w = 15\ nm$. The Power was kept constan for all measurements. From fitting $\eta_{meas} = A_0 + \eta_0 \cdot e^{-L/L_m}$, the following values were extracted for $\eta_{meas}(IL= 0) = A_0 + \eta_0$: **(a)** $A_0$ = 0 % + 1.6 % , $\eta_0$ = 8.2 % ± 1.6 %, $L_m$ = 12.3 μm ± 3.6 μm; **(b)** $A_0$ = 2.6 % ± 1.1 %, $\eta_0$ = 5.6 % ± 1.0 %, $L_m$ = 10.5 μm ± 3.0 μm; **(c)** $A_0$ = 2.8 % ± 1.2 %, $\eta_0$ = 5.8 % ± 1.0 %,, $L_m$ = 7.1 μm ± 2.7 μm; **(d)** $A_0$ = 3.7 % ± 0.34 %, $\eta_0$ = 4.5 % ± 0.3 %, , $L_m$ = 4.4 μm ± 0.8 μm; **(e)** $A_0$ = 1.91 % ± 0.6 %, $\eta_0$ = 6.3 % ± 1.1 %, $L_m$ = 4.1 μm ± 1.7 μm.

The propagation length of the HGPW mode for varying gap width, $w$, can be evaluated via the so-called *cut-back* method[5]. The in- and out-coupled signal of the entire system is compared for waveguides of constant widths $w$ but different lengths $L$, as shown in Figure S3. The detected signal, $\eta_{meas}$, for the different waveguide lengths can be approximated by an exponential decay law of the plasmonic mode in the waveguide, $\eta_{meas} \propto \eta_0 e^{-L/L_m}$ with the initial amplitude being $\eta_0$ and the fundamental mode's propagation length $L_m$. This method possesses the great advantage that $L_m$ can be determined independently of the respective taper and grating efficiencies. The extracted propagation lengths as a function of gap width are shown and compared to theory in Figure S4a.

We define $\eta_{meas}$ as the ratio between the in- and out-coupled signal via the gratings, measured using a Peltier-cooled IR camera at constant integration time, integration area and pump power of 6.6 µW. The incident light, coupled to the waveguide via the in-coupling grating, and from the waveguide to the out-coupling grating was evaluated via the signal, $S_{cts,out}$, emitted from the grating. This value is normalised by a reference signal, $R_{cts,in}$, collected in reflection from thick, fully reflecting Au patch over the same area and at the same power:

$$\eta_{meas} = \frac{S_{cts,out} - B_{cts,out}}{R_{cts,in} - B_{cts,in}} . \tag{S2.1}$$

The signal and pump counts have been corrected by the background signals, $B_{cts,out}$ and $B_{cts,in}$, respectively. During the measurement, an iris positioned in the image plane after the objective (c.f. Figure S11) allowed us to isolate the signal from the out-coupling grating spatially from the rest of the sample.

The efficiency measurements shown in Figure S3 provide an estimate for the in-coupling efficiency $\eta_{inc,gap}$ into the plasmonic gap waveguide at ($L = 0$). This has been extracted from the measurements using:

$$\eta_{meas}(l = 0) = \eta_{inc,gap}\eta_{outc,gap}\eta_{Gr-Gr} \Leftrightarrow \eta_{inc,gap} = \sqrt{\frac{\eta_{meas}(L = 0)}{1.1 \cdot \eta_{Gr-Gr}}} , \tag{S2.2}$$

where $\eta_{outc,gap}$ is the out-coupling efficiency. $\eta_{Gr-Gr} \approx 9\%$ is the grating-to-grating coupling efficiency, which we were able to measure directly using a control sample without the metal taper shown in Figure S1c. At first glance, one might assume that the focusing and de-focusing efficiencies would be approximately the same, $\eta_{inc,gap} \approx \eta_{outc,gap}$. We have checked this by simulations (3D FDTD) to find that $\eta_{outc,gap}/\eta_{inc,gap} = 1.1$, which is the value used in Eq. (S2.2). The focusing process is sensitive to the launching of the correct wavefronts at the in-coupling grating, whereas the de-focusing process is not. We expect that a perfectly implemented coupling grating could ensure focusing and de-focusing with the same efficiency. The values for $\eta_{meas}(L=0)$ at each measured gap width are listed in the caption of Figure S3.

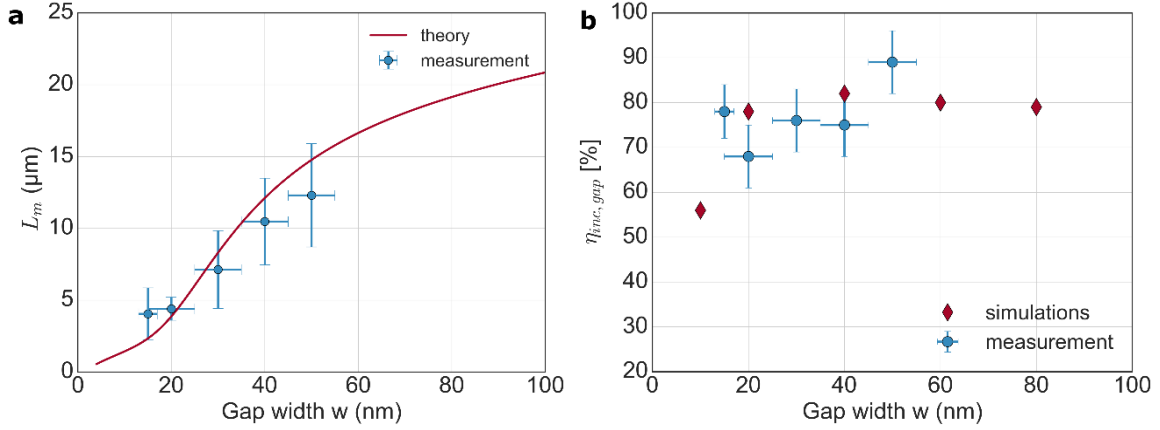

**Figure S4|** **a** Extracted propagation length from the cut-back method measurements shown in Figure S3 for various gap widths, compared to theoretical eigenmode-solver calculations. **b** Extracted gap in-/out-coupling efficiencies compared to 3D FDTD simulations of the entire system at a wavelength of 1550 nm.

Figure S4b compares the measured (extracted from Figure S3) and calculated coupling efficiencies. The latter has been determined from 3D FDTD simulations[6]. The result demonstrates a good agreement between experiment and theory, which underpins the good performance of this nanofocusing platform with a gap coupling efficiency of $\eta_{inc,gap} \approx 80\ \%$. Based on this result, the out-coupling efficiency of erbium fluorescence (out of the gap and guided to the grating) has been conservatively estimated to be 80 %. This is likely to be closer to 90 %, consistent with simulations; however, this cannot be verified experimentally.

## 4. Rate equation model and the Purcell factor $F_P$ in saturation regime

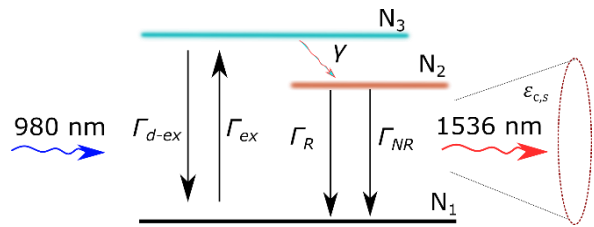

**Figure S5|** **Schematic three-level energy diagram illustrating the main Er$^{3+}$-ion 4f populated states,** (with population numbers $N_1$, $N_2$, $N_3$) and corresponding transitions probed in the experiment. The 980 nm pump and the 1536 nm emission signal (peak wavelength) are shown in blue and red, respectively. $\Gamma_{ex}$ and $\Gamma_{d-ex}$ are the excitation and de-excitation rates between $N_3$ and $N_1$. $\gamma$ is the transition rate from level 3 to level 2. $\Gamma_R$ and $\Gamma_{NR}$ describe the radiative and non-radiative transitions from level 2 to 1. $\varepsilon_{c,s}$ is the collection efficiency of light emitted by the transition from level 2 to level 1.

The measured fluorescence signal $S_S$ of an ion coupled to an optical mode (either photonic or plasmonic) can be expressed by the radiative de-population rate $\Gamma_R$ of energy level 2 with population $N_2$ (c.f. Fig. S4) and the emission collection efficiency $\varepsilon_{c,s}$:

$$S_S = \varepsilon_{c,s} N_2 \Gamma_R \ . \tag{S3.1}$$

As the population $N_2$ and its radiative de-population rate $\Gamma_R$ are not directly accessible in this experiment, we will use a three-level rate equation model to find an experimentally accessible expression. The excitation and recombination dynamics illustrated in Figure S5 can be described by Einstein's rate equations for a three-level system:

$$\dot{N}_2 = -\Gamma_R N_2 - \Gamma_{NR} N_2 + \gamma N_3 \tag{S3.2}$$

$$\dot{N}_3 = (N_1 - N_3)\Gamma_{ex} - \gamma N_3 \tag{S3.3}$$

$$N = N_1 + N_2 + N_3, \tag{S3.4}$$

where $N$ is the total population number of the system, $\gamma$ is the transition rate of carriers from level 3 to level 2 with populations numbers $N_3$ and $N_2$, $\Gamma_{NR}$ is the non-radiative recombination rate, and $\Gamma_{ex}$ is the excitation rate of carriers from the ground state to level 3. Under steady state illumination, we assume that the excitation rate is proportional to optical pump power, $\Gamma_{ex} \propto P$.

For CW excitation we assume a steady state solution: $\dot{N}_2 = \dot{N}_3 = 0$. For rapid de-population of the third level, $N_3 \approx 0 \Rightarrow N_1 \approx N - N_2$ [7,8], we can write:

$$N_2 \Gamma_R = \frac{N\Gamma_{ex}}{[\Gamma_{ex} + \Gamma_R + \Gamma_{NR}]} \tag{S3.5}$$

Substitution into Eq. (S3.1) yields:

$$S_s = \frac{\varepsilon_{c,s} N \Gamma_{ex} \eta_{q,s}}{\left[1 + \frac{\Gamma_{ex}}{\Gamma}\right]}, \tag{S3.6}$$

where $\Gamma = \Gamma_R + \Gamma_{NR}$ is the total de-population rate, which takes the radiative and non-radiative recombination into account. Accordingly, the internal quantum efficiency $\eta_{q,s} = \Gamma_R/\Gamma$ is identified.

For our Gr-Gr configuration control sample, the measured signal, $S_{Gr}$, can be expressed in analogue to Eq. (S3.6):

$$S_{Gr} = \varepsilon_{c,Gr} \widetilde{N}_{Gr} \Gamma_R \frac{\Gamma_{ex,Gr}/\Gamma}{\left[1 + \frac{\Gamma_{ex,Gr}}{\Gamma}\right]}, \tag{S3.7}$$

where the total population number $\widetilde{N}_{Gr}$ is directly proportional with a constant, $a$, to the number of illuminated ions $\widetilde{N}_{Gr} = aN_{Gr}$ in the Gr-Gr configuration. $\varepsilon_{c,Gr}$ is the photon coupling efficiency to the detector, $\Gamma_{ex,Gr}$ the excitation rate from the ground level, and $\eta_{q,Gr}$ the internal quantum efficiency as defined for the general case in Eq. (S3.6).

Let us now take the Purcell effect into account to describe the accelerated emission of the hybrid plasmonic gap waveguide devices. The Purcell enhancement factor $F_p$ contains radiative and non-radiative components and may be expressed as $F_p = F_{pR} + F_{pNR}$. However, $F_p$ acts only upon the radiative process, with recombination rate $\Gamma_R$, given that the non-radiative rate $\Gamma_{NR}$ is due to internal processes of the atomic system and is not affected by altering the electromagnetic density of states. Hence, a modified total de-population rate $\Gamma'$ and modified internal quantum efficiency $\eta_{q,Gap}$ are defined:

$$\Gamma \to \Gamma' = F_p\Gamma_R + \Gamma_{NR}, \qquad (S3.8)$$

$$\eta_{q,s} \to \eta_{q,Gap} = \frac{F_p\Gamma_R}{[F_p\Gamma_R+\Gamma_{NR}]}. \qquad (S3.9)$$

Analogously to Eq. (S3.7), we can write the measured signal of Er$^{3+}$-ion emitters in the hybrid plasmonic gap waveguide devices using the substitutions ($\Gamma'$ and $\eta_{q,Gap}$),

$$S_{Gap} = \varepsilon_{c,Gap}\widetilde{N}_{Gap}F_p\Gamma_R \frac{\Gamma_{ex,Gap}/\Gamma'}{\left[1+\frac{\Gamma_{ex,Gap}}{\Gamma'}\right]}, \qquad (S3.10)$$

where the total population $\widetilde{N}_{Gap}$ is directly proportional to the number of illuminated gap ions $\widetilde{N}_{Gap} = aN_{Gap}$ in the experiment. $\varepsilon_{c,Gap}$ is the photon coupling efficiency to the detector, $\Gamma_{ex,Gap}$ the excitation rate from the ground level, and $\eta_{q,Gap}$ the internal quantum efficiency as defined for the general case in Eq. (S3.6).

Next, it is insightful to consider the low pump power and the high pump power limits for $S_{Gr}$ and $S_{Gap}$. Firstly, let us consider the low pump power case in which the excitation rate is much lower than the radiative recombination rate. Considering the non-plasmonic case $S_{Gr}$ and where the pump power, $P \propto \Gamma_{ex,Gr} \ll \Gamma = \Gamma_R + \Gamma_{NR}$, we can write:

$$S_{Gr} \approx \varepsilon_{c,Gr}\widetilde{N}_{Gr}\Gamma_{ex,Gr}\eta_{q,Gr}. \qquad (S3.11)$$

Similarly, in the plasmonic case $S_{Gap}$ and assuming $P \propto \Gamma_{ex,Gap} \ll \Gamma' = F_P\Gamma_R + \Gamma_{NR}$, yields:

$$S_{Gap} \approx \varepsilon_{c,Gap}\widetilde{N}_{Gap}\Gamma_{ex,Gap}\eta_{q,Gap}. \qquad (S3.12)$$

Comparing the ratio of detected signals for Purcell-enhanced and the non-enhanced devices:

$$\frac{S_{Gap}}{S_{Gr}} = F_p \frac{\varepsilon_{c,Gap}}{\varepsilon_{c,Gr}} \frac{\widetilde{N}_{Gap}}{\widetilde{N}_{Gr}} \frac{\Gamma_{ex,Gap}}{\Gamma_{ex,Gr}} \frac{\eta_{q,Gap}}{\eta_{q,Gr}} \qquad (S3.13)$$

The Purcell factor cannot be reliably determined from this expression since the excitation rates and quantum efficiencies cannot easily be extracted from measurements.

In the high pump power case, $P \propto \Gamma_{ex,Gr} \gg \Gamma = \Gamma_R + \Gamma_{NR}$, leaving us with:

$$S_{Gr} \approx \varepsilon_{c,Gr}\widetilde{N}_{Gr}\Gamma_R. \qquad (S3.14)$$

For the Gr-Gr device and:

$$S_{Gap} \approx \varepsilon_{c,Gap}F_p\widetilde{N}_{Gap}\Gamma_R \qquad (S3.15)$$

For the hybrid gap plasmon waveguide devices. The ratio of these saturated signals yields an expression for the Purcell factor $F_p$:

$$\frac{S_{Gap}}{S_{Gr}} = \frac{\varepsilon_{c,Gap}}{\varepsilon_{c,Gr}} \frac{N_{Gr}}{N_{Gap}} F_p .  \quad (S3.16)$$

## 5. Gaussian saturation model and Purcell enhancement results

In the following, we use the fact that the excitation rate $\Gamma_{ex}$ is linearly proportional to the pump beam's intensity, $\mathcal{I}(x,y)$, i.e., $\Gamma_{ex} = \gamma_{ex}\mathcal{I}(x,y)$. We cannot follow exactly the saturation model from Section 3, due to the Gaussian excitation beam distribution. Following Eq. (S3.6) we can write the output intensity from the 3-Level model for the Gr-Gr case as,

$$I = c\varepsilon_C\Gamma_R \frac{\mathcal{I}(x,y)/\mathcal{I}_{sat}}{1+\mathcal{I}(x,y)/\mathcal{I}_{sat}}, \quad (S4.1)$$

where $\mathcal{I}_{sat} = \Gamma/\gamma_{ex}$ is the saturation intensity, $c$ is the Erbium ion areal density, $\varepsilon_C$ is the coupling efficiency of ion emission to the detector, and $\Gamma_R$ is the recombination rate of Erbium ions. For a Gaussian beam, $\mathcal{I}(x,y) = ae^{-2r^2/r_0^2}$, we find a relationship with excitation power, $P = \pi r_0^2 a/2$, so that,

$$I = c\varepsilon_C\Gamma_R \frac{e^{-2r^2/r_0^2}P/P_{sat}}{1+e^{-2r^2/r_0^2}P/P_{sat}} \quad (S4.2)$$

where, $\mathcal{I}_{sat}\pi r_0^2/2 = P_{sat}$, is the saturation power. To determine the total collected signal, $S(P)$, that would be measured in an experiment, we integrate over all excited ions,

$$S(P) = c\varepsilon_C\Gamma_R \int_0^{2\pi} d\phi \int_0^\infty r dr \frac{e^{-2r^2/r_0^2}P/P_{sat}}{1+e^{-2r^2/r_0^2}P/P_{sat}}. \quad (S4.3)$$

Here we have assumed that the collection efficiency, areal density, and emission rate of the excited ions are uniform across the illumination area. To simplify, we distinguish the saturation function from the saturation signal by substituting for a normalised integration variable: $\rho = \sqrt{2}r/r_0$, to find:

$$S(P) = \frac{1}{2}r_0^2 c\varepsilon_C\Gamma_R \int_0^{2\pi} d\phi \int_0^\infty \rho d\rho \frac{e^{-\rho^2}P/P_{sat}}{1+e^{-\rho^2}P/P_{sat}} \quad (S4.4)$$

We use this saturation model to evaluate the saturation of ions in the gap of the plasmonic waveguide (HGPW device) and the SiO$_2$ on SOI slab waveguide surface (Gr-Gr device).

**Saturation Model for the Gr-Gr Control Sample:**

For the Gr-Gr control sample, we need only to evaluate the integral, to find,

$$S_{Gr}(P) = \frac{1}{2}A_{Gr}c\varepsilon_{C,Gr}\Gamma_R \ln(1 + P/P_{Gr,sat}) = S_{Gr,sat}\ln(1 + P/P_{Gr,sat}) \quad (S4.5)$$

where $A_{Gr} = \pi r_0^2$, is the area of the Gaussian beam and $\varepsilon_{C,Gr}$ is the collection efficiency of Erbium emission for the Gr-Gr sample. This equation can be fit to the control sample data using two free parameters, $S_{Gr,sat}$ and $P_{Gr,sat}$. For the computation of the Purcell factor, we will want to find a value for $S_{Gr,sat} = A_{Gr} c \varepsilon_{C,Gr} \Gamma_R / 2$, from the fitted experimental data.

**Saturation Model for the Gap Plasmon Samples:**

For the gap plasmon samples, Erbium emission originates from the gap and taper regions, which we treat separately. For the taper region, we assume that the Erbium emission is related to that of the Gr-Gr control sample but limited by the physical masking of the Erbium by the metal structure that defines the taper. For a taper opening angle $\alpha$ as defined in the main text, we find,

$$S_{taper}(P) = \alpha r_0^2 c \varepsilon_{C,Gr} \Gamma_R \int_0^\infty \rho d\rho \frac{e^{-\rho^2} P/P_{Gr,sat}}{1+e^{-\rho^2} P/P_{Gr,sat}} = \frac{\alpha}{\pi} S_{Gr}(P) \tag{S4.6}$$

For the Erbium ions in the gap, we assume a near uniform excitation profile ($e^{-2r^2/r_0^2} \approx 1$) to produce the conventional saturation response function:

$$S_{Gap}(P) \approx c \varepsilon_C F_p \Gamma_R \frac{P/P_{Gap,sat}}{1+P/P_{Gap,sat}} \int_{-w/2}^{w/2} dy \int_{-L/2}^{L/2} dx = A_{gap} c \varepsilon_{C,Gap} \Gamma_R F_p \frac{P/P_{Gap,sat}}{1+P/P_{Gap,sat}}, \tag{S4.7}$$

where $A_{gap} = wL$ is the area of the gap region, $\varepsilon_{C,Gap}$ is the collection efficiency of ions from the gap region, and $F_P$ is the Purcell factor. The total saturation response of the Gap sample is now described by $S_h(P) = S_{Gap}(P) + S_{taper}(P)$, such that

$$S_h(P) \approx S_{Gap,sat} \frac{P/P_{Gap,sat}}{1+P/P_{Gap,sat}} + \frac{\alpha}{\pi} S_{Gr,sat} \ln(1 + P/P_{Gr,sat}) \tag{S4.8}$$

where $S_{Gap,sat} = A_{gap} c \varepsilon_{C,Gap} \Gamma_R F_p$ and $S_{Gr,sat} = A_0 c \varepsilon_{C,Gr} \Gamma_R / 2$. Here again, only two fitting parameters are used, $S_{Gap,sat}$ and $P_{Gap,sat}$, as $S_{Gr,sat}$ and $P_{Gr,sat}$ are taken from the model fitting (Eq. (S4.5)) to the Gr-Gr sample data.

**Evaluating the Purcell Factor from Saturation Curves**

Experimental data is thus fit to the expressions for the Gr-Gr control samples, $S_{Gr}(P)$, and the gap plasmon devices, $S_h(P)$. The Purcell factor can now be computed using the ratio:

$$F_p = \frac{S_{Gap,sat}}{S_{Gr,sat}} \frac{A_{Gr}}{A_{gap}} \frac{\varepsilon_{C,Gr}}{\varepsilon_{C,Gap}} \equiv \langle F_p \rangle \tag{S4.9}$$

The parameters extracted from fitting Eq. (S4.5) and Eq. (S4.8) to the measured integrated fluorescence as a function of power and for different gap widths, shown in Figure 2c of the main manuscript, are listed in Table S1. In the experiment, the collected signal of all illuminated ions can contribute to the total detected signal. The individual ion Purcell enhancement however is dependent on its position with respect to the gap. Thus we determine an average Purcell factor $\langle F_p \rangle$ based on the measurement.

**Table S1| Fit parameters and reduced $\chi^2_{red}$**, extracted from least square optimization of the fit functions for $S_i(P)$ with $i = \{Gr, Gap\}$, describing the power dependency of the integrated measured emission spectra in Gr-Gr and HGPW configuration for different gap widths $w$.

| Gap width $w$ | $S_{i,sat}$ (photons/s) | $P_{i,sat}$ (mW) | $\chi^2_{red}$ |
|---|---|---|---|
| 10 nm | $(80 \pm 11) * 10^4$ | $72 \pm 14$ | 1.31 |
| 17 nm | $(59 \pm 16) * 10^4$ | $91 \pm 30$ | 0.45 |
| 30 nm | $(47 \pm 14) * 10^4$ | $113 \pm 44$ | 1.03 |
| 60 nm | $(22 \pm 4) * 10^4$ | $58 \pm 15$ | 3.21 |
| Gr-Gr | $(10 \pm 1) * 10^4$ | $23 \pm 4$ | 1.01 |

**Table S2| Measured average Purcell factor $\langle F_p \rangle$ and the simulated radiative Purcell factor $\langle F_{p,1nm} \rangle$ averaged across the lateral dipole positions of the gap waveguide at 1 nm below the interlayer oxide/gap interface.** The values are listed for various gap widths $w$. $\varepsilon_{c,Gr}$ and $\varepsilon_{c,Gap}$ are the collection efficiency via the gratings and the fundamental gap mode, respectively. $\frac{A_{Gr}}{L \cdot w}$ is the ratio between the illumination area in the Gr-Gr configuration, $A_{Gr}$ and the area of the waveguide gap $A_{Gap} = w\,L$ of length $L$ and width $w$. All values are for an emission wavelength of λ = 1.536 μm.

| Gap width $w$ | $\varepsilon_{c,Gr}/\varepsilon_{c,Gap}$ | $\frac{A_{Gr}}{A_{Gap}}$ | $\langle F_p \rangle$ - measured | $\langle F_{p,1nm} \rangle$ - simulated |
|---|---|---|---|---|
| 10 nm | 0.23 | $317 \pm 100$ | $592 \pm 220$ | $600 \pm 182$ |
| 17 nm | 0.23 | $187 \pm 38$ | $256 \pm 91$ | $224 \pm 57$ |
| 30 nm | 0.23 | $106 \pm 15$ | $116 \pm 42$ | $103 \pm 54$ |
| 60 nm | 0.23 | $53 \pm 6$ | $27 \pm 7$ | $62 \pm 50$ |

Table S2 lists the average Purcell enhancement $\langle F_p \rangle$ extracted from measurement and simulations. In simulations, we average the purely radiative Purcell factor $F_p$ of a single dipole emitter over different positions across the gap for one specific implantation depth Δd (Figure S6). The simulations are performed for an electric dipole oriented perpendicular to the gap waveguide direction (TE) positioned at a distance of Δd below the Au/SiO$_2$ interface. Table S2 list the result for Δd = 1 nm. The simulation results of individual Purcell factor, $F_p$, are shown exemplarily below for the Δd = 1 nm, Δd

= 4 nm and Δd = 8 nm as described in the main text for 1 nm and 4 nm (Fig. 3a). The simulated $\langle F_p \rangle$ neglects the $F_p$ values for dipoles with a distance of < 1 nm from the metal wall as these are strongly quenched in the experiment[9]. Note that the values for $F_p$ shown in Figure S6 are normalized by the Purcell enhancement due to vicinity of the SOI substrate which accounts for a factor of ≈ 1.8.

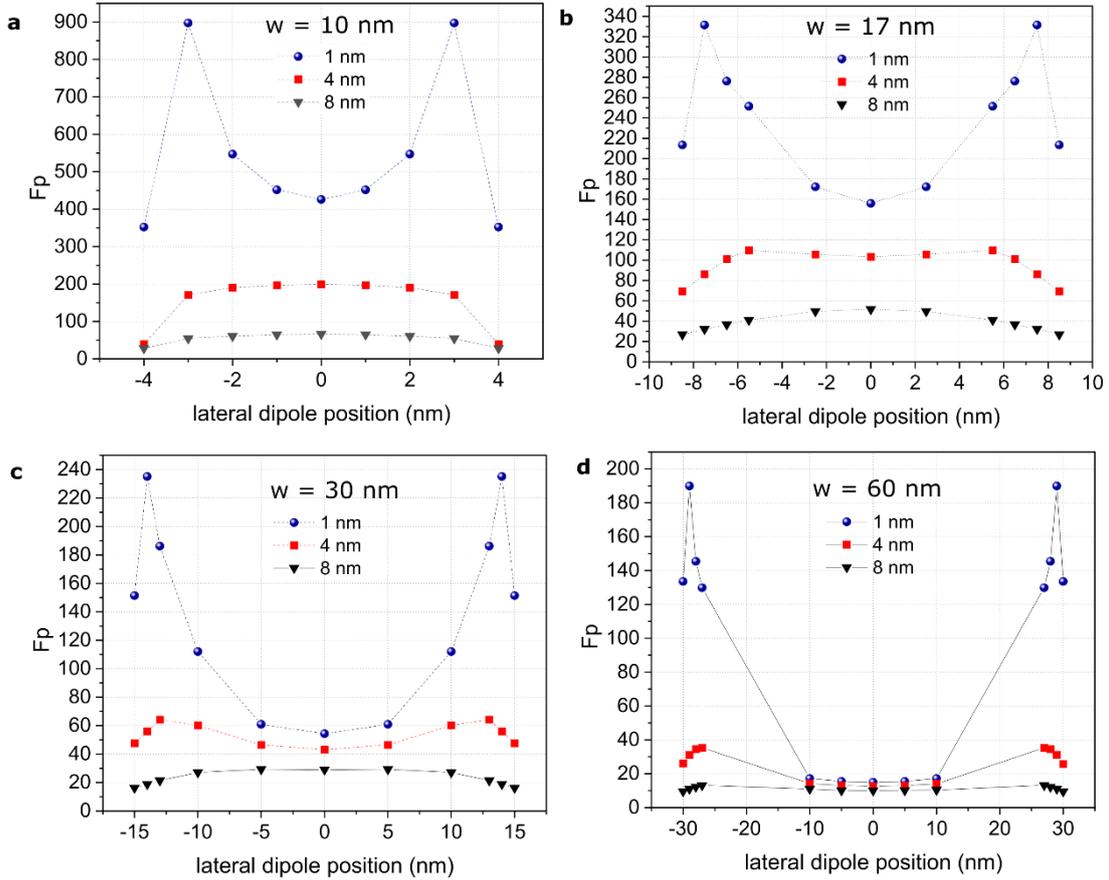

**Figure S6|** FDTD simulations showing the radiative Purcell enhancement $F_p$ for a perpendicular to the MIM gap aligned electrical dipole at different distances, Δd = {1 nm, 4 nm, 8 nm} below the SiO$_2$/Au gap interface for various lateral positions (centre gap position at 0 nm). Radiative Purcell enhancement for MIM gap widths of **a** 10 nm, **b** 17 nm, **c** 30 nm and **d** 60 nm all for a wavelength of 1.536 µm and a 50 nm thick Au gap layer.

## 6. Polarization dependence of pump beam to waveguide coupling

The pump beam, which excites the erbium ions in the metal-insulator-metal (MIM) gap from the top, is linearly polarized. The coupling of pump power into the gap strongly depends on the pump beam's polarization axis with respect to the waveguide's orientation, i.e. perpendicular to the gap vs. parallel to the gap. This is due to the fact that the fundamental mode of the MIM structure under study is transverse electric (TE), with field lines perpendicular to the longitudinal orientation of the gap (c.f. Figure S7a1. Hence, solely TE (perpendicular) polarized light couples to the fundamental gap mode as illustrated in Figure S7. This has been studied in the literature[10,11] and is specifically shown here for a $w = 10$ nm gap waveguide in the structure studied in the main manuscript.

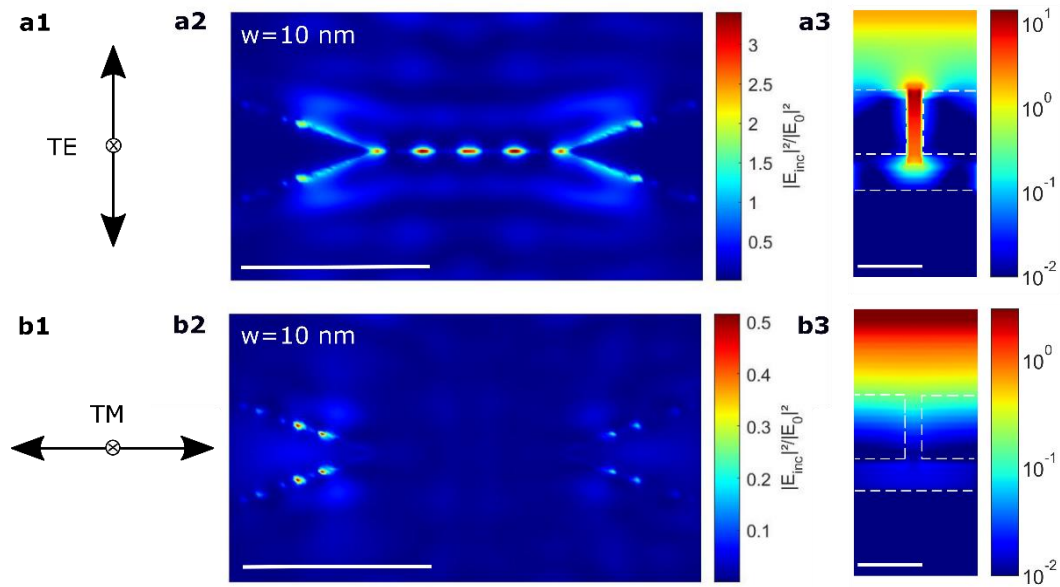

**Figure S7 | Polarization dependence of a 10 nm wide gap waveguide at 980 nm pump wavelength.** **a1**, TE polarization of the pump beam with a perpendicular field direction with respect to the longitudinal gap. **a2**, 3D FDTD simulation showing the normalized field intensity cross-section $|E_{inc}|^2/|E_0|^2$ at a depth of 1 nm underneath the gap waveguide for a waveguide system with $w = 10\ nm$ gap width (horizontal monitor). Scale bar 1 μm. **a3**, 3D FDTD simulation showing the normalized field intensity cross-section $|E_{inc}|^2/|E_0|^2$ (log-scale) as a side view of the gap waveguide system for a 10 nm gap (vertical monitor). Scale bar 50 nm, the white dotted lines indicate the contours of the Au gap and the Si-slab substrate, respectively. The spacer interlayer is $SiO_2$. **b1 - b3**, same simulations as in **(a2)** and **(a3)** but with a TM polarized pump beam. All intensity values are normalized with respect to the same incident pump intensity.

Figure S7a2 and b2 show 3D FDTD simulations of the electrical field intensity at the position of the ions, 1 nm underneath the waveguide for TE and TM, respectively. Weak backscattering at both ends of the waveguide results in a lateral eigenmode resonance (standing wave pattern) of the waveguide with a field enhancement of about 3. This can be only observed for the TE case, in stark contrast to the TM case for which the field intensity in the gap is negligible. This is underpinned by the vertical cross-sections shown in Figure S7a3 and Figure S7b3, illustrating the electric field intensity in the gap waveguide for TE and TM pump polarization, respectively. The ratio of pump field intensity in the gap for the two polarizations varies by about two orders of magnitude. Thus, rotating the pump polarization relative to the gap, essentially allows the fluorescence from ions in the gap to be switched on and off.

The dependence of the gap in-coupling efficiency on the pump polarization is not only shown in simulations (shown in Figure S7) but can also be observed in the experiment. Figure S8 shows the dependence of the integrated emission signal on pump beam polarization at different positions on the sample and its comparison to the expected $\cos^2(\varphi)$ law, where $\varphi$ is the polarization angle. The pump wavelength is 980 nm, the same as in the measurements in the main text. A pump polarization of 0° corresponds to a polarization parallel to the gap (i.e. TM) and a pump polarization of 90° corresponds to a polarization perpendicular to the gap (i.e. TE).

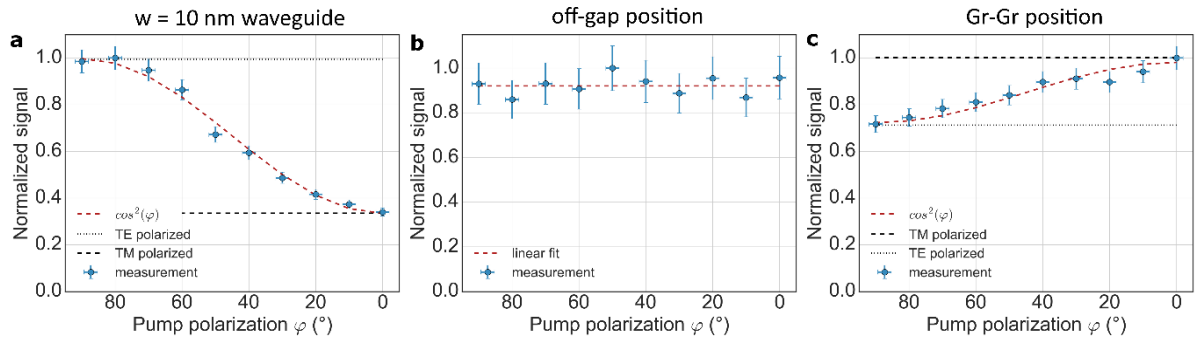

**Figure S8| Pump beam polarization (φ) measurements of integrated emission for different configurations at a wavelength of 980 nm. a** Central illumination of the HGPW structure with a $w = 10\ nm$ gap waveguide. **b** off-gap position for reference (no gratings and no gap). **c** Central illumination of the grating-to-grating (Gr-Gr) configuration (no gap waveguide). The signal at a pump beam polarization perpendicular to the gap is labelled "TE polarized" and the signal at parallel polarization of the pump beam is labelled "TM polarized". Error bars were estimated to 5% for the waveguide and the Gr-Gr configuration and 10% of the off-gap position based on the standard deviation at the same power from the main manuscript.

Figure S8a shows the polarization dependence of the normalized signal, and hence the in-coupled light into the gap, for a $w = 10$ nm hybrid gap waveguide. One observes that the measured signal drops by about 66% when a TM instead of a TE pump polarization is used. In accordance with the simulations shown above, we conclude that at least 66% of the collected signal originates from the ions illuminated in the gap waveguide alone. The rest of the collected signal stems from non-enhanced ions from outside of the waveguide gap. This is further underpinned by that fact that the same measurement at a position away from the plasmonic waveguide shows very little pump polarization dependence, as illustrated in Figure S8b. Moreover, the measurement in Figure S8c shows that 66% is a lower estimate on the proportion of gap fluorescence, since in absence of the gap waveguide (i.e. the laser spot is centred between two focused gratings in "Gr-Gr'' configuration) the TM polarized pump beam couples $\approx 30\%$ more efficiently to the gratings compared to the TE polarized pump. Thus, we estimate a signal contribution of erbium ions in the gap region to the entire detected signal in the HGPW configuration to be larger than 78% ($\approx 66\% + 30\% * 40\%$).

## 7. Signal collection efficiency

This section explains how the collection efficiency ratio $\frac{\varepsilon_{c,Gr}}{\varepsilon_{c,Gap}}$ was estimated. Figure S9 shows a simplified illustration of the emission pathways of the ion fluorescence in the HGPW and the Gr-Gr configuration. The emitter is excited by linear polarized light (TE, perpendicular to the gap). The emission collected in the experiment can be estimated by considering two major pathways: i) fraction of light vertically radiated i.e. in normal direction off the sample ($X_{rad,a}$ in gap configuration and $X_{rad,b}$ in Gr-Gr) where it is collected by an objective with a solid angle of ratio $\Omega_r$; ii) fraction of light coupled to the TE fundamental mode in both configurations ($X_{TE,a}$ for HGPW and $X_{TE,b}$ for Gr-Gr) which propagates to gratings of collection efficiency $\eta_{Gr}$. The collection efficiency ratio of the two configurations can be written as:

$$\frac{\varepsilon_{c,Gr}}{\varepsilon_{c,Gap}} = \frac{X_{TE,b}\,\eta_{Gr}\,\Omega_{Gr}+X_{rad,b}\,\Omega_r}{X_{TE,a}\,\eta_{Gr}+\Omega_r\,X_{rad,a}} \qquad (S6.1)$$

where the collection of light emitted by a horizontal dipole which is parallel to the pump polarization is estimated to $\Omega_{Gr} = \frac{50°}{120°} \approx 42\%$, based on a 50° opening angle of each grating for the Gr-Gr configuration and the dipole emission angle in plane of around 120° as an upper estimate; a lower estimate for the coupling efficiency of ion emission from the gap to the TE slab mode (entirely collected by the gratings) is $X_{TE,a} \approx 80\%$, extracted from experiments outlined in Figure S4b, as the focused grating couplers are designed to collect all the light exiting the gap, which however can be subject to slight reflections during the out-coupling from the gap; fraction of light coupled to the TE slab mode from the dipole $X_{TE,b} \approx 0.5 * 60\% = 30\%$ where a TE/TM mode coupling ratio of 50:50 to the slab was conservatively estimated and 60% of dipole emission coupled to the slab, extracted from simulations; grating coupling efficiency $\eta_{Gr} \approx 30\%$ extracted from the experiment; the percentage of vertically emitted light from ions placed in a plain $SiO_2$ on Si system, 1 nm – 8 nm below the interface as described in the main text, is $X_{rad,b} \approx 20\%$ as extracted from simulations; the percentage of vertically emitted light out of the gap towards the objective for different gap width $w = [10\ nm, 17\ nm, 30\ nm, 60\ nm]$ is $X_{rad,b}$= [0.03%, 0.04%, 0.24%, 0.24% ]; The percentage of collected light by the objective (NA=0.4) used in the experiment is determined by the ratio of solid angles while for $X_{rad,a}$ and $X_{rad,b}$ the simulation settings where such that approximately the entire emission in a half sphere was collected, resulting in $\Omega_r = A = sin^2\left(\frac{\theta_1}{2}\right) * (sin^2(\frac{\theta_0}{2}))^{-1}$ = 0.083, with $\theta_1 = 23.58°$ and $\theta_0 = 90°$; finally, we arrive at a lower bound estimate for the collection ratio of $\frac{\varepsilon_{C,Gr}}{\varepsilon_{C,Gap}} \geq 23\%$ based on Eq. (S6.1), which shows only little dependence on the gap width.

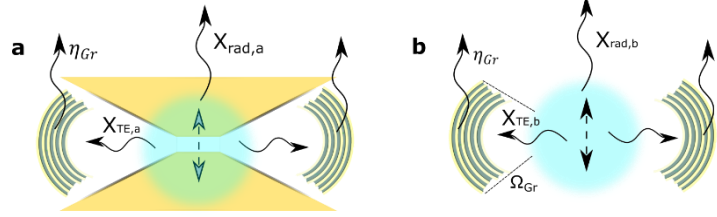

**Figure S9|** Sketch, illustrating the emission pathways for in this case a polarized excited emitter (linear double arrow) in (a) the HGPW configuration and (b) the Gr-Gr setting.

## 8. Purcell enhancement waveguide coupling of electric-dipole (ED) and magnetic-dipole (MD) emission

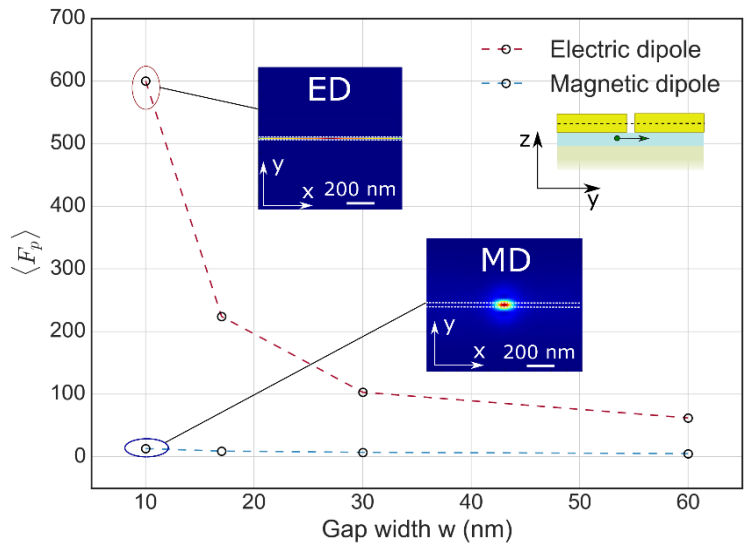

**Figure S10| Simulated average Purcell factor $\langle F_p \rangle$ for an electric- and magnetic dipole emitter centrally positioned at $\Delta d = 1\ nm$ below the plasmonic nanogap at λ = 1.536 μm.** The upper right inset shows the cross-section of the metal-insulator-metal (MIM) gap waveguide where the arrow indicates the dipole orientation. Here, the dashed black line indicates the position of the monitor which records the top view cross-sections of electric field $|E|$ due to the ED and MD coupling to the gap, shown on the left and on the bottom, respectively (both for $w = 10\ nm$, white dotted lines mark the gap edges). The MD emission enhancement appears to be negligible in comparison to the ED emission enhancement and the MD emission does not couple efficiently to the predominantly transverse electric MIM gap mode.

## 9. Experimental setup

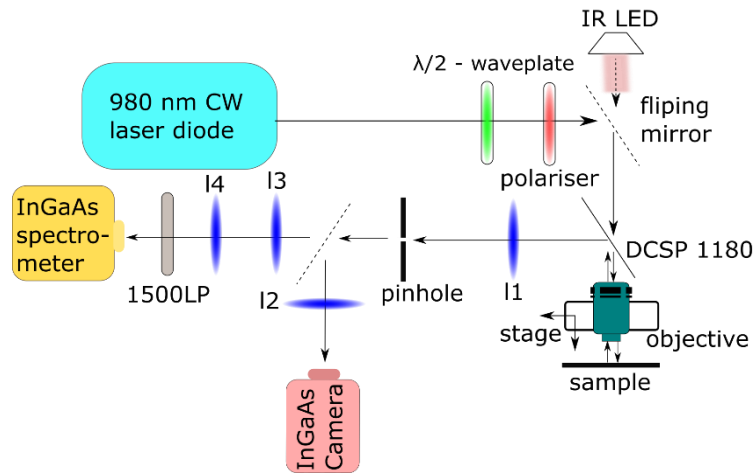

**Figure S11| Schematic of the experimental beamline** using an Agilent FPL4916 CW laser diode, a Mitutoyo infinity-corrected achromatic near-IR objective (NA=0.4, 20x, 2 cm WD) and a Princeton Instruments (ActonSP2300) nitrogen cooled (−120°C) InGaAs 1D CCD detector array. For the spectrometer grating, a groove density of 600 gr/mm and a blaze wavelength of 1.6 µm with an efficiency of ≈85% (manufacturer information) at 1536 nm has been used. The lenses l1 - l4 collimate the beam onto the IR camera, the pinhole position, and the spectrometer. A dichroic short pass filter centred at 1180 nm (DCSP 1180) has been used, to allow the 980 nm CW pump to illuminate the sample while collecting the reflected signal from the sample at around 1536 nm. The pinhole/iris merely serves to isolate the signal of a single coupling grating from the rest of the sample.

## 10. Fluorescence lifetime measurements

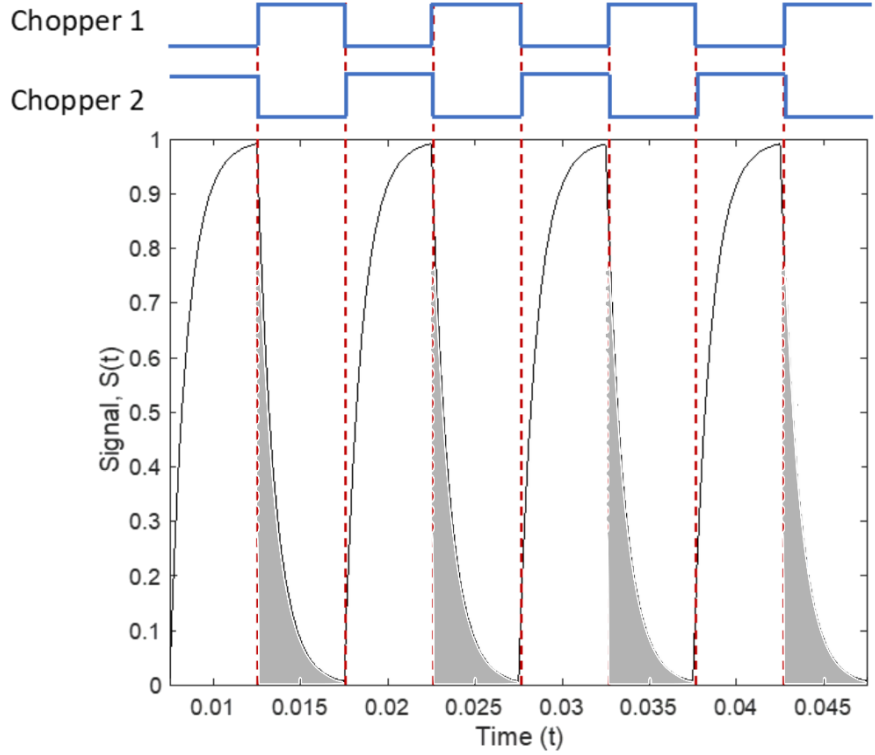

**Figure S12| Illustration of square-wave modulated signal, $S(t)/S_0$, as a function of time for a modulation frequency of $\nu = 100$ Hz.** The modulation of the pump beam (chopper 1) and detected signal (chopper 2) are illustrated as $\pi$ out of phase. The shaded areas indicate where the detector integrates the signal leading to a value of $P_\pi/\nu t_{int}$ for each period.

The lifetimes of erbium ions implanted within the Gr-Gr control and 10 nm HGPW devices were measured using a modulated pump and detection technique. Both the pump light and the fluorescence were focused through mechanical choppers to produce near-square-wave responses, as shown in Figure S12. The excited state lifetime of erbium ions for square wave excitation at a frequency $\nu$ over a single period of modulation, $T = \nu^{-1}$, can be described by a three-level model. Assuming rapid transfer of electrons from the excited $^4I_{11/2}$ state to the fluorescent $^4I_{13/2}$ state, and a low pump rate to avoid state saturation, a single differential equation may be used to describe the $^4I_{13/2}$ population,

$$\frac{dN}{dt} = -\frac{N}{\tau} + R(t)N_T, \qquad (S9.1)$$

where $R = r$ for $0 < t < T/2$, and $R = 0$ for $T/2 < t < T$. The emitted signal under continuous excitation is $S_0 = N/\tau = rN_T$. Under a modulated signal, it can be shown that, $S(t) = N(t)/\tau$,

$$S(t) = S_0\left(1 - \frac{e^{-t/\tau}}{(1+e^{-T/2\tau})}\right), \quad 0 < t < T/2 \qquad (S9.2)$$

$$S(t) = S_0\frac{e^{-(t-T/2)/\tau}}{(1+e^{-T/2\tau})}, \quad T/2 < t < T, \qquad (S9.3)$$

and this response is illustrated in Figure S12. A second chopper, phase-locked to the first, enables the signal to be measured over an integration time, $t_{int}$, with a phase shift, $\phi$, relative to the first modulation. This enables the lifetime to be spectrally resolved by using an imaging spectrometer. For a modulation frequency, $\nu$, the detector integrates over $\nu t_{int}$ periods. Assuming an experimental collection and detection signal of $\eta$, for out-of phase detection ($\phi = \pi$) we measure a signal,

$$P_\pi(\nu) = \eta \nu t_{int} S_0 \int_{T/2}^{T} \frac{e^{-(t-T/2)/\tau}}{(1+e^{-T/2\tau})} dt = \eta \nu \tau t_{int} S_0 \tanh((4\nu\tau)^{-1}). \tag{S9.4}$$

To normalize this signal, we measure the sample fluorescence without the two choppers over the same integration time, to find $P_* = \eta t_{int} S_0$. To determine the lifetime, we thus calculate

$$F(\nu) = \frac{P_\pi(\nu)}{P_*} = \nu\tau \tanh((4\nu\tau)^{-1}). \tag{S9.5}$$

For a collection of emitters with a single exponential decay, the lifetime is returned for each measurement of $F(\nu)$. In the case of multi-exponential decay, the frequency resolved response is necessary to correctly distinguish the various decay components. Note that for $\nu\tau \ll 1$, $\tanh((4\nu\tau)^{-1}) \approx 1$, a linear variation is found, $\nu\tau \approx F(\nu)$, which provides a fixed frequency link to the lifetime.

**Multi-exponential decay**

For mixtures of emitters with differing lifetimes, we will find an average signal,

$$P_\pi(\nu) = P_* \sum_i^N a_i \nu\tau_i \tanh((4\nu\tau_i)^{-1}) \tag{S9.6}$$

where $a_i$ are the proportions of emitters with lifetimes, $\tau_i$, and $\sum_i a_i = 1$. In the bi-exponential case with $a_2 = 1 - a_1$

$$P_\pi(\nu) = P_*\nu[a_1\tau_1\tanh((4\nu\tau_1)^{-1}) + (1-a_1)\tau_2\tanh((4\nu\tau_2)^{-1})]. \tag{S9.7}$$

This is the situation observed in our experiments. There are thus three parameters to find to determine the variation of $P_\pi(\nu)/P_*$. Note also that there are three regimes, which provide information about the mixture of emitters and their lifetimes:

| $\nu\tau_1, \nu\tau_2 \ll 1$ | $\nu\tau_1 > 1, \nu\tau_2 \ll 1$ | $\nu\tau_1, \nu\tau_2 > 1$ |
|---|---|---|
| $\dfrac{P_\pi(\nu)}{P_*} = \nu(a_1\tau_1 + a_2\tau_2)$ | $\dfrac{P_\pi(\nu)}{P_*} = \dfrac{a_1}{4} + (1-a_1)\nu\tau_2$ | $\dfrac{P_\pi(\nu)}{P_*} = \dfrac{1}{4}$ |

These formulae explain the data observed in experiments on the Gr-Gr control and 10 nm gap HGPW samples described in the main text. Figure S13 shows the experimental data with the theoretical fits to these data.

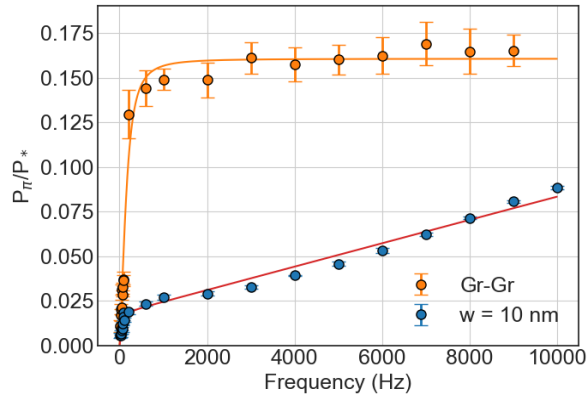

**Figure S13| Frequency dependency of the ratio of modulated to unmodulated power detected for the Gr-Gr and 10 nm gap devices.** Both devices were excited with a 980 nm CW laser. Solid lines are the fitted bi-exponential decay functions Eq. (S8.7).

To cover the span of the modulation frequency range, a dual frequency blade and a chopper with 100 slots were used. The dual frequency blade was used to generate the low modulation frequency from 2 Hz to 100 Hz, while the second chopper was used to generate high modulation frequency from 200 Hz to 10 kHz.

**Experimental parameters:**

The fluorescence of the Gr-Gr control device without modulation $P_*$ was measured, by exciting the sample with a 980 nm CW laser at the power of 40 mW over the integration time of 30 seconds. The out of phase fluorescence signal $P_\pi$ was measured by using the same pump power and integration with both choppers on. The HGPW sample was measured by using a lower pump power of 20 mW. The other conditions were identical to the Gr-Gr control device measurement.

**Fitting parameters:**

The signal $P_\pi(\nu)/P_*$ as a function of the modulation frequency $\nu$ was fitted by using the bi-exponential decay function

$$P_\pi(\nu) = P_*\nu[a_1\tau_1\tanh((4\nu\tau_1)^{-1}) + (1-a_1)\tau_2\tanh((4\nu\tau_2)^{-1})]. \qquad (S9.8)$$

The fitting parameters of the Gr-Gr sample are $a_1 = 0.643$, $\tau_1 = 1$ ms, $\tau_2 = 5$ ns.

The parameters of the 10 nm gap sample are $a_1 = 0.071$, $\tau_1 = 2.5$ ms, $\tau_2 = 7.1$ μs.

**Measuring the power dependence of lifetime.**

Generally, the lifetime will depend on the pump power. Hence, the natural lifetime results from extrapolation to the value at zero-pump power. Once the modulation frequency dependence of the emitters was determined, the lifetime could be extracted at a fixed modulation frequency as a function of the pump power, $r$. For a single exponential decay, we can use the following relationship:

$$\nu \frac{P_*(r)}{P_\pi(r)} \approx \tau(r)^{-1}. \tag{S9.9}$$

For a bi-exponential decay, the same approach can be used. For the slower emitters, $\nu\tau_1 \ll 1$, $\tau_1 \gg \tau_2$,

$$\nu a_1 \frac{P_*(r)}{P_\pi(r)} \approx \tau_1(r)^{-1}. \tag{S9.10}$$

The power dependence of the faster emitters in the regime $\nu\tau_1 > 1, \nu\tau_2 \ll 1$, can also be determined using the expression:

$$\frac{(1-a_1)\nu}{\frac{P_\pi(r)}{P_*(r)} - \frac{a_1}{4}} = \tau_2(r)^{-1}. \tag{S9.11}$$

Such measurements were conducted on the Gr-Gr and 10 nm gap HGPW devices at fixed frequency as a function of pump power, as shown in Figure 4c of the main text. This data shows a linear dependency of the emission rates on pump power.

**Spectrally resolved lifetime**

This technique allows rapid collection of spectrally resolved lifetime data by using an imaging spectrometer. Figure 14 shows the spectrally resolved lifetime of erbium ions from the Gr-Gr sample ($\Delta\lambda = 20$ nm binning). Due to the weak signal from this sample, the spectral resolution is limited. Nonetheless, the erbium lifetime is uniform across the erbium emission band with an average lifetime of 2.622 ms ± 0.718 ms. The spectrally resolved enhancement factor (lifetime ratio) for the 10 nm gap HGPW sample is shown in Figure 4d of the main text.

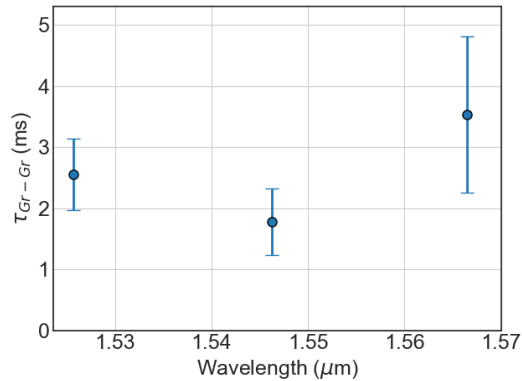

**Figure 14| Spectrally resolved lifetime of erbium ions from the Gr-Gr sample.** The lifetime is extracted from the power dependent emission rate measurements, with excitation power from 5 $mW$ to 40 $mW$. To ensure good signal to noise ratio, spectral resolution of the lifetime is set to be around 20 nm.

## 11. Power dependence of the signal spectrum

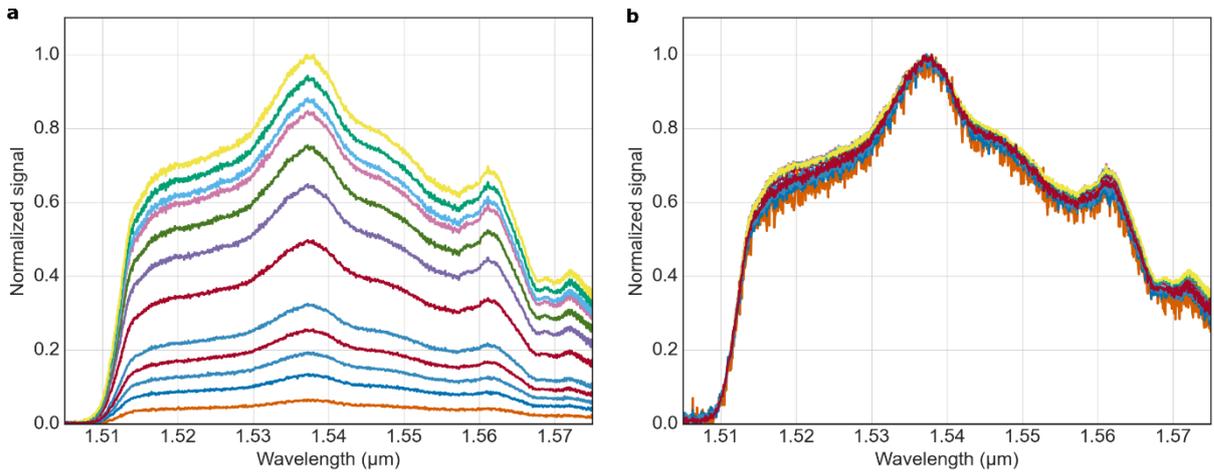

**Figure S15| Normalized fluorescence spectra** measured at the pump power $[2, 4, 6, 8, 10, 15, 20, 25, 30, 35, 40]\ mW$ (from orange to yellow color-coding) for a waveguide structure with a $w = 10\ nm$ gap. **a**, Normalized to the overall maximum signal and **b**, normalized to the maximum signal of each spectrum. No pump power-dependent red-shift or change of the spectral shape is observed, which excludes any form of self-stimulated emission.

## 12. Fluorescence background

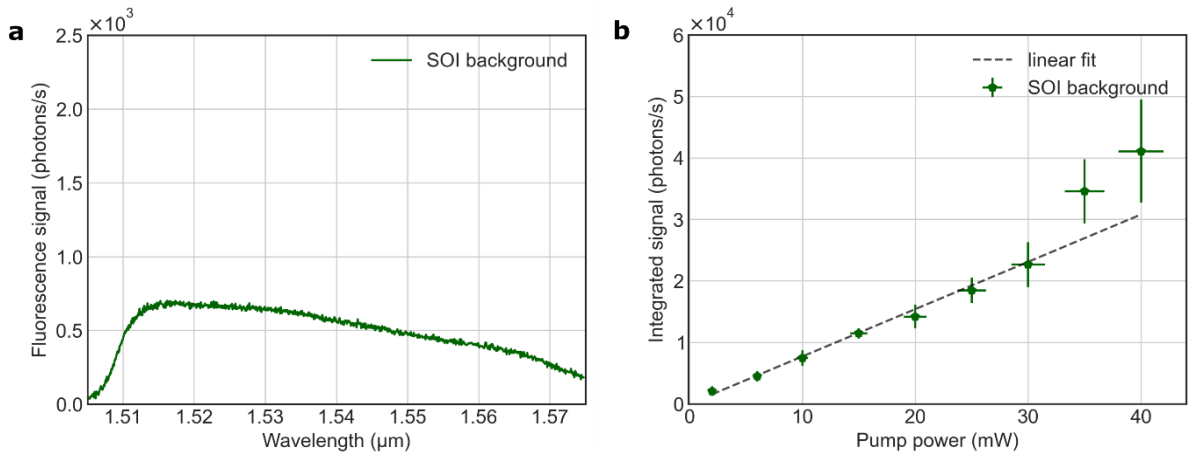

**Figure S16| Background fluorescence from a non $Er^{3+}$-ion containing Si reference sample. a**, Fluorescence spectra from a no ion containing reference sample measured at 40 mW pump power (980 nm, CW). The sample was annealed analogously to the actual devices used in the main manuscript. **b**, Power dependent measurement showing the integrated signal from **(a)** at different pump powers. Error bars are the standard deviation. The dashed lines indicate the theory fits to each curve. Linear fitting was performed to estimate the Si background subtracted from the measurement in the manuscript.

To account for background fluorescence from the sample, reference measurements were performed on a plain SOI sample without Erbium ions. The sample was coated with 25 nm $SiO_2$, identical to the ion containing sample. The "SOI background" reference sample spectra is shown in Figure S16a. This sample was annealed under the same conditions as the ion containing sample described in the main manuscript but does not possess any patterns or fabricated structures. To provide a conservative overall estimate of the Purcell enhancement factor, $\langle F_p \rangle$, we assumed a linear power dependence of the SOI background (Figure S16b). The slight non-linear trend may arise from a nonlinear behaviour of the photodetector for the very low count rates of the non-ion containing samples. For the estimate of $I_{i,sat}$ and eventually $\langle F_p \rangle$ in the main manuscript, the linear fitted power dependence of the background was subtracted from the measured fluorescence power dependencies of the ion-containing samples. In the case of the HGPW devices, the background signal was multiplied by the fraction of the illuminated sample area which is not covered by Au before subtraction. The background subtraction is used for the curves in the fluorescence power dependencies shown in Figure 2c.